\documentclass[twocolumn]{aastex631}

\shorttitle{A volatile-rich atmosphere on WASP-121b}
\shortauthors{Pelletier et al.}

\begin{document}

\title{\large{CRIRES$^{+}$ and ESPRESSO reveal an atmosphere enriched in volatiles relative to refractories on the ultra-hot Jupiter WASP-121b}}

\author[0000-0002-8573-805X]{Stefan Pelletier}
\affiliation{Department of Physics and Trottier Institute for Research on Exoplanets, Universit\'{e} de Montr\'{e}al, Montreal, QC, Canada}
\affiliation{Observatoire astronomique de l'Université de Genève, 51 chemin Pegasi 1290 Versoix, Switzerland}

\author[0000-0001-5578-1498]{Bj\"orn Benneke}
\affiliation{Department of Physics and Trottier Institute for Research on Exoplanets, Universit\'{e} de Montr\'{e}al, Montreal, QC, Canada}

\author[0000-0003-1728-8269]{Yayaati Chachan}
\altaffiliation{CITA National Fellow}
\affiliation{Department of Physics, McGill University, Montreal, QC H3A 2T8, Canada}
\affiliation{Department of Astronomy and Astrophysics, University of California, Santa Cruz, CA 95064, USA}

\author[0000-0003-3181-5264]{Luc Bazinet}
\affiliation{Department of Physics and Trottier Institute for Research on Exoplanets, Universit\'{e} de Montr\'{e}al, Montreal, QC, Canada}

\author[0000-0002-1199-9759]{Romain Allart}
\altaffiliation{Trottier Fellow}
\affiliation{Department of Physics and Trottier Institute for Research on Exoplanets, Universit\'{e} de Montr\'{e}al, Montreal, QC, Canada}

\author[0000-0001-8981-6759]{H.\ Jens Hoeijmakers}
\affiliation{Lund Observatory, Department of Astronomy and Theoretical Physics, Lund University, Box 43, 221 00 Lund, Sweden}

\author[0000-0001-8477-5265]{Alexis Lavail}
\affiliation{Institut de Recherche en Astrophysique et Planétologie, Université de Toulouse, CNRS, IRAP/UMR5277, Toulouse, France}

\author[0000-0001-7216-4846]{Bibiana Prinoth}
\affiliation{Lund Observatory, Department of Astronomy and Theoretical Physics, Lund University, Box 43, 221 00 Lund, Sweden}
\affiliation{European Southern Observatory, Alonso de C\'ordova 3107, Vitacura, Regi\'on Metropolitana, Chile}

\author[0000-0002-2195-735X]{Louis-Philippe Coulombe}
\affiliation{Department of Physics and Trottier Institute for Research on Exoplanets, Universit\'{e} de Montr\'{e}al, Montreal, QC, Canada}

\author[0000-0003-3667-8633]{Joshua D.\ Lothringer}
\affiliation{Department of Physics, Utah Valley University, Orem, UT 84058, USA}
\affiliation{Space Telescope Science Institute, Baltimore, MD 21218, USA}

\author[0000-0001-9521-6258]{Vivien Parmentier}
\affiliation{Observatoire de la C\^ote d'Azur, CNRS UMR 7293, BP4229, Laboratoire Lagrange, F-06304 Nice Cedex 4, France}

\author[0000-0002-9946-5259]{Peter Smith}
\affiliation{ School of Earth and Space Exploration, Arizona State University, Tempe, AZ 85287, USA}

\author[00000-0002-4085-6001]{Nicholas Borsato}
\affiliation{School of Mathematical and Physical Sciences, Macquarie University, Sydney, NSW 2109, Australia}
\affiliation{Lund Observatory, Department of Astronomy and Theoretical Physics, Lund University, Box 43, 221 00 Lund, Sweden}

\author[0000-0002-5633-4400]{Brian Thorsbro}
\affiliation{Observatoire de la C\^ote d'Azur, CNRS UMR 7293, BP4229, Laboratoire Lagrange, F-06304 Nice Cedex 4, France}
\affiliation{Department of Astronomy, School of Science, The University of Tokyo, Tokyo, Japan}



\begin{abstract}

One of the outstanding goals of the planetary science community is to measure the present-day atmospheric composition of planets and link this back to formation. 
As giant planets are formed by accreting gas, ices, and rocks, constraining the relative amounts of these components is critical to understand their formation and evolution. 
For most known planets, including the Solar System giants, this is difficult as they reside in a temperature regime where only volatile elements (e.g., C, O) can be measured, while refractories (e.g., Fe, Ni) are condensed to deep layers of the atmosphere where they cannot be remotely probed.
With temperatures allowing for even rock-forming elements to be in the gas phase, ultra-hot Jupiter atmospheres provide a unique opportunity to simultaneously probe the volatile and refractory content of giant planets.
Here we directly measure and obtain bounded constraints on the abundances of volatile C and O as well as refractory Fe and Ni on the ultra-hot giant exoplanet WASP-121b. We find that ice-forming elements are comparatively enriched relative to rock-forming elements, potentially indicating that WASP-121b formed in a volatile-rich environment much farther away from the star than where it is currently located. 
The simultaneous constraint of ice and rock elements in the atmosphere of WASP-121b provides insights into the composition of giant planets otherwise unattainable from Solar System observations.

\end{abstract}



\section{Introduction} \label{sec:introduction}


The atmospheres of planets can offer us a window into their compositions.  
But often, this only gives us a glimpse - a small piece of the full picture.  
Such is the case for the Solar System gas giants where, despite dedicated satellite missions \citep{niemann_galileo_1992,fletcher_methane_2009,janssen_mwr_2017}, we are only able to measure the abundance of highly or moderately volatile elements like C, N, O, S, and the noble gases \citep{mahaffy_noble_2000,wong_updated_2004,li_water_2020, li_super-adiabatic_2024}. 
Meanwhile, no direct measurements exist for any refractory elements (e.g., Fe, Mg, Si, Ni) due to the freezing cold temperatures ($<200$\,K) on Jupiter and Saturn, leading to all of the rock-forming elements to condense to the deep interior where they cannot be directly measured \citep{lewis_clouds_1969}. 
This makes constraining the ice-to-rock ratio by directly measuring both the refractory and volatile compositions effectively impossible for most known planets.  
Even for Jupiter, Saturn, Uranus, and Neptune, the current uncertainty in the relative amounts of rocky and icy material accreted during formation hinders our understanding of how and where in the protosolar disk the planets in our own Solar System formed~\citep{mousis_determination_2009, mousis_nature_2021, thiabaud_elemental_2015, venturini_jupiters_2020}.

The advent of exoplanets has allowed us to finally study the planets in our Solar System as a population rather than a single data point. Even on worlds that can be hundreds of light years away, their hotter temperatures often mean that volatile molecules such as H$_2$O, CO, CO$_2$, and CH$_4$ are well-amenable to remote sensing with current instrumentation~\citep[e.g.,][]{barman_identification_2007, snellen_orbital_2010, birkby_detection_2013, bell_methane_2023, welbanks_high_2024}.
This has motivated numerous works studying how different C and O abundances on exoplanet atmospheres (often parameterized as metallicity and C/O ratio) can arise from different formation conditions~\citep{oberg_effects_2011, madhusudhan_co_2012, madhusudhan_toward_2014, madhusudhan_atmospheric_2017, eistrup_setting_2016, eistrup_molecular_2018, cridland_connecting_2019, cridland_connecting_2020, schneider_how_2021}.  Carbon and oxygen are of particularly high interest in the context of planet formation as the two most abundant elements in the universe after hydrogen and helium and the main components of icy material in protoplanetary discs.
However, C and O abundances alone cannot uniquely determine where in the protoplanetary disk a giant planet accreted its envelope~\citep{mordasini_imprint_2016, khorshid_simab_2022, molliere_interpreting_2022, pacetti_chemical_2022}.  
The degeneracy arises in part because these atoms can be held up and accreted in both solid or gaseous form depending on the local disk conditions.  For example, O held in H$_2$O will be vaporized in the gas phase in the warm inner disk regions, but condensed into solid ice beyond the water snowline~\citep[e.g.,][]{stevenson_rapid_1988}.

Refractory elements, on the other hand, always remain in solid form throughout all but the inner-most hottest regions of the disc~\citep{lodders_solar_2003}.  
Hence, the abundance of refractories on a planet can act as a direct measure of the fractional amount of solid material accreted, independent of formation location, and therefore help break the degeneracies that arise from considering only C and O~\citep{turrini_tracing_2021, chachan_breaking_2023, fonte_oxygen_2023, crossfield_volatile--sulfur_2023}.  Measuring the relative abundances of refractory and volatile elements in the atmosphere of a giant planet can therefore track the rock-to-ice ratio of accreted solids~\citep{lothringer_new_2021, schneider_how_2021-1}. However, most planets reside in temperature regimes where the principal refractory building blocks of planet formation are condensed out of the gas-phase and effectively impossible to directly probe via remote sensing.

Ultra-hot Jupiters present an unprecedented opportunity to simultaneously probe both volatile and refractory elements in giant planetary atmospheres~\citep[e.g.,][]{lothringer_new_2021, kasper_unifying_2022, ramkumar_high-resolution_2023}.  
Specifically, we argue that ultra-hot Jupiters in the $T_{\mathrm{eq}} \sim$ 2,200 -- 2,400\,K range are ideal targets for accurately constraining volatile-to-refractory abundance ratios.  In this regime, the temperature is hot enough that rock-forming elements like Fe are in gaseous form, but not so hot that metals and molecules become significantly ionized or dissociated~\citep{kitzmann_peculiar_2018, lothringer_extremely_2018, parmentier_thermal_2018, landman_detection_2021, nugroho_first_2021, brogi_roasting_2023}.
Ultra-hot Jupiters also generally have puffier atmospheres and more favorable flux contrasts relative to their colder counterparts, and are expected to have cloud-free daysides~\citep{helling_cloud_2021}, making them ideal target for atmospheric characterization.
While more moderately refractory species such as Na, K, Li, or S can also be probed in the atmosphere of lower temperature planets~\citep[e.g.,][]{wyttenbach_spectrally_2015, casasayas-barris_detection_2017, seidel_hot_2019, welbanks_massmetallicity_2019, borsa_atmospheric_2021, tsai_photochemically_2023, benneke_jwst_2024, sing_warm_2024, welbanks_high_2024}, these elements may not be fully condensed in the protoplanetary disk and thus could only represent a lower limit of the amount of solids accreted~\citep{chachan_breaking_2023}.

While ultra-hot Jupiters are expected to have both rock-forming species and volatile molecules in gaseous form in their dayside atmosphere, measuring these simultaneously to constrain key elemental abundance ratios such as O/Fe and C/Fe remains notoriously challenging.  The difficulty arises because the primary refractory and volatile components of ultra-hot Jupiters have prominent spectral features in different wavelength regimes. Atomic metals like Fe or Ni (refractories) generally have spectral lines that are strongest in the optical and weaker in the infrared.  Contrastingly, molecules such as CO and H$_2$O (volatiles) tend to be relatively transparent at optical wavelengths (hence why we can see stars in the sky on a clear dark night), with more prominent opacity in the infrared. Accessing a wide spectral coverage at high spectral resolution can therefore be a powerful approach for measuring both refractory metals and volatile molecules, although this is not always strictly necessary~\citep[e.g.,][]{ramkumar_high-resolution_2023, parker_into_2024}.

We further note that while JWST is currently unmatched at detecting molecules that have broad and distinct spectral bands (e.g., H$_2$O, CH$_4$, CO$_2$, SO$_2$), species such as CO and Fe that have strong but narrow absorption lines can be more difficult to unambiguously probe with lower spectral resolving powers~\citep{spake_abundance_2021, lothringer_new_2021, bell_methane_2023, benneke_jwst_2024}.  
Meanwhile, ground-based high-resolution spectroscopy can act as a powerful complementary method to probe absorbers that lack strong distinct bandheads~\citep[e.g.,][]{brogi_carbon_2014, brogi_exoplanet_2018, hoeijmakers_atomic_2018, hoeijmakers_spectral_2019, hoeijmakers_high-resolution_2020, nugroho_searching_2020, giacobbe_five_2021, carleo_gaps_2022, guilluy_gaps_2022, kasper_unifying_2022, prinoth_titanium_2022, kesseli_atomic_2022, borsato_mantis_2023, smith_combined_2024, bazinet_subsolar_2024}.

\begin{figure*}[t]
\begin{center}
\includegraphics[width=\linewidth]{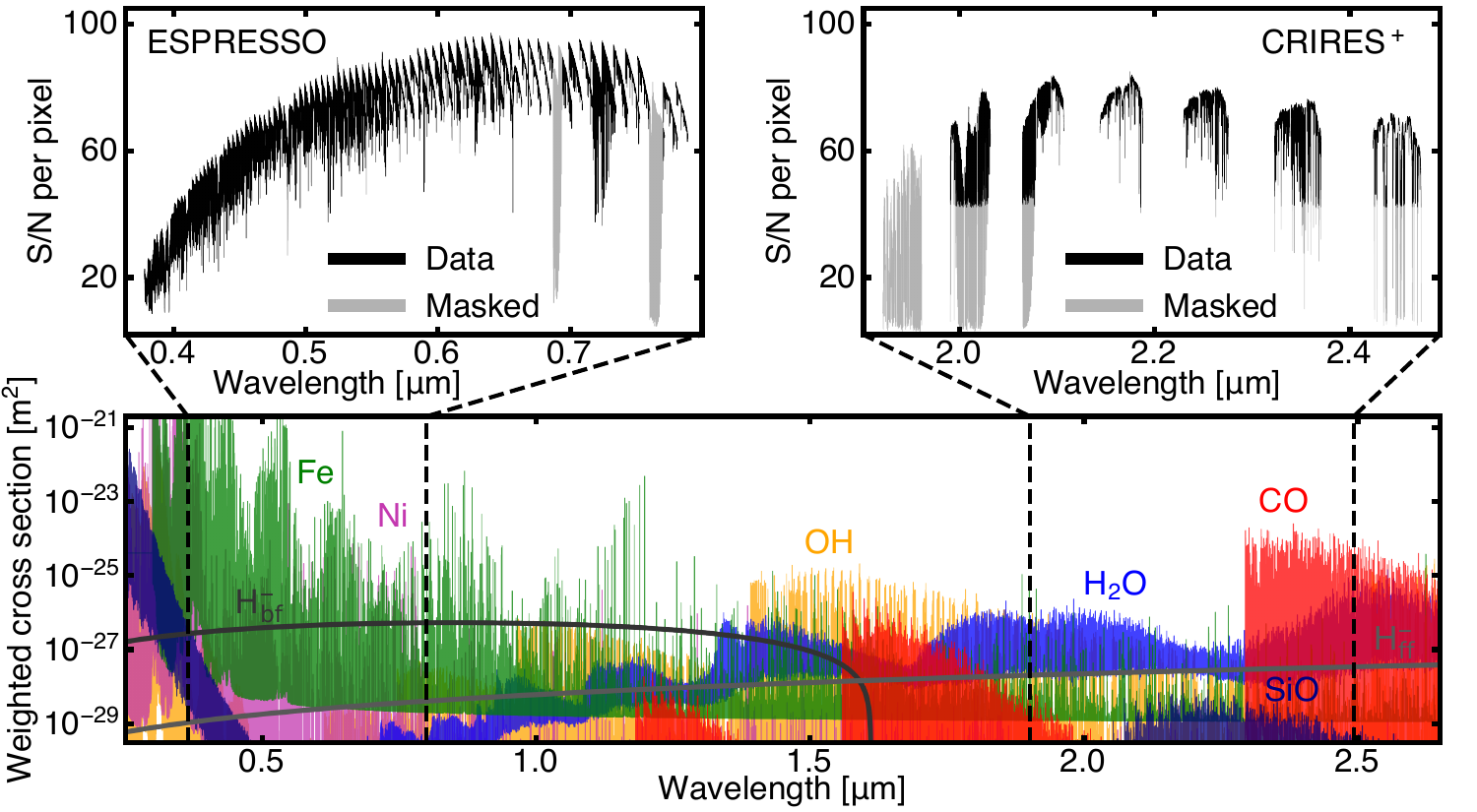}
\end{center}
\vspace{-4mm}
\caption{Example ESPRESSO and CRIRES$^{+}$ exposures compared to cross sections of important refractory, volatile, and continuum species.  
The top panels show the signal-to-noise per pixel of a typical ESPRESSO (left) and CRIRES$^{+}$ (right) exposure, with the regions of the data that are masked out during the analysis shown in gray. The bottom panel shows weighted cross sections of refractory elements Fe and Ni, volatile molecules H$_2$O, CO, OH, and SiO, and continuum opacity sources H$^{-}_{\mathrm{bf}}$ and H$^{-}_{\mathrm{ff}}$ computed at a temperature of 2900\,K. 
The black dashed lines show the wavelength bounds of the ESPRESSO (0.38 -- 0.78\,$\mu$m) and CRIRES$^{+}$ K2166 (1.92 -- 2.48\,$\mu$m) data used in work.
ESPRESSO is well suited for probing Fe and Ni, while CRIRES$^{+}$ is sensitive to CO and H$_2$O.  However, CRIRES$^{+}$ in the $K$-band mode is not particularly sensitive to other important O-bearing molecules such as OH or SiO.  Combining observations in the optical and near-infrared can be a powerful method to constrain volatile-to-refractory abundance ratios.
}
\label{fig:cross_sections}
\end{figure*}

In this work we present high-resolution observations of the dayside atmosphere of WASP-121b using both CRIRES$^{+}$ and ESPRESSO.  
In Section~\ref{sec:observations} we present the observations.  The methods of our analysis are outlined in Section~\ref{sec:methods}.  Cross-correlation and retrieval results are shown in Section~\ref{sec:results} and their implications in the context of planet formation are discussed in Section~\ref{sec:formation}.  Finally, we conclude in Section~\ref{sec:conclusion}.


\section{Observations} \label{sec:observations}
We observed the dayside of the ultra-hot Jupiter WASP-121b ($R_p = 1.753\pm0.036$\,$R_{\rm Jup}$, $M_p = 1.157\pm0.070$\,$M_{\rm Jup}$, $P = 1.2749250(1)$\,days, $T_{\mathrm{eq}} = 2358\pm52$\,K)~\citep{delrez_wasp-121_2016, bourrier_hot_2020, patel_empirical_2022} using the high-resolution CRIRES$^{+}$ infrared slit spectrograph~\citep{follert_crires_2014, dorn_crires_2023} at the European Southern Observatory Very Large Telescope in Paranal, Chile.  We targeted WASP-121b at orbital positions before secondary eclipse on 13 February 2022, and after secondary eclipse on 14 December 2021, each for a duration of 2.8 hours. While photometric for the pre-eclipse observations, the weather conditions were less favorable during the post-eclipse observations. 
The exposures were taken in an ABBA nodding sequence (alternating between two positions on the slit) to best remove background sky emission, with each detector integration lasting 300\,s.  The observations made use of the adaptive optics system, and were done in the $K$-band K2166 wavelength setting.

For the CRIRES$^{+}$ data, we used the default ESO pipeline, \texttt{EsoRex}, to extract wavelength calibrated spectra from each observed exposure by running the \texttt{cr2res\_obs\_nodding} recipe.  For each night, we compute the spectral resolution from the median full width at half maximum (FWHM) of the point spread function (PSF).  We measure this to be about $R$ = 120,000 for the first observing night and $R$ = 105,000 for the second observing night.  Although different for the two nights, the spectral resolutions were stable to within 1-2\% over the course of both observing sequences.  The spectral resolutions being above the expected $R$ = 100,000 resolution of CRIRES$^{+}$ using the 0.2'' slit are a results of the high performance of the adaptive optics causing the point spread function to be smaller than the slit width~\citep{dorn_crires_2023, yan_crires_2023, lesjak_retrieval_2023}. We perform our analysis on the outputted combined AB spectra comprising of seven wavelength segments between 1.92 and 2.48\,$\mu$m.  A typical extracted CRIRES$^{+}$ spectrum of WASP-121 is shown in Figure~\ref{fig:cross_sections} (top right panel).

We supplement our CRIRES$^{+}$ observations of the dayside atmosphere of WASP-121b with the ESPRESSO data from \cite{hoeijmakers_mantis_2024}. ESPRESSO~\citep{pepe_espresso_2021} is a high-resolution optical spectrograph covering the 0.38--0.78\,$\mu$m wavelength range at a resolution of $R \sim $ 140,000.  The ESPRESSO observation of WASP-121b consist of eight epochs, each of a 3--4\,h duration, evenly split targeting pre- and post-eclipse orbital phases. A typical extracted ESPRESSO spectrum of WASP-121 is shown in Figure~\ref{fig:cross_sections} (top left panel).  We refer the reader to \cite{hoeijmakers_mantis_2024} for more detail regarding the ESPRESSO data.

The motivation for combining CRIRES$^{+}$ and ESPRESSO is that volatile molecules such as H$_2$O and CO have strong opacity in the infrared while refractory elements have their strongest spectral lines predominantly at visible/UV wavelengths (Figure~\ref{fig:cross_sections}, bottom panel).  Combining CRIRES$^{+}$ with ESPRESSO thus permits us to probe both volatile and refractory species and constrain the relative proportion of ices and rocks in the atmosphere of WASP-121b.


\section{Methods} \label{sec:methods}
We perform cross-correlation and retrieval analyses on the combined ESPRESSO and CRIRES$^{+}$ data to characterize the chemistry and thermal structure of the dayside atmosphere of WASP-121b.  The overall analysis procedure closely follows \cite{pelletier_vanadium_2023} and is briefly summarized here.  

\subsection{Cleaning the data of telluric and stellar contributions} \label{subsec:detrending}

\begin{figure*}[t]
\begin{center}
\includegraphics[width=\linewidth]{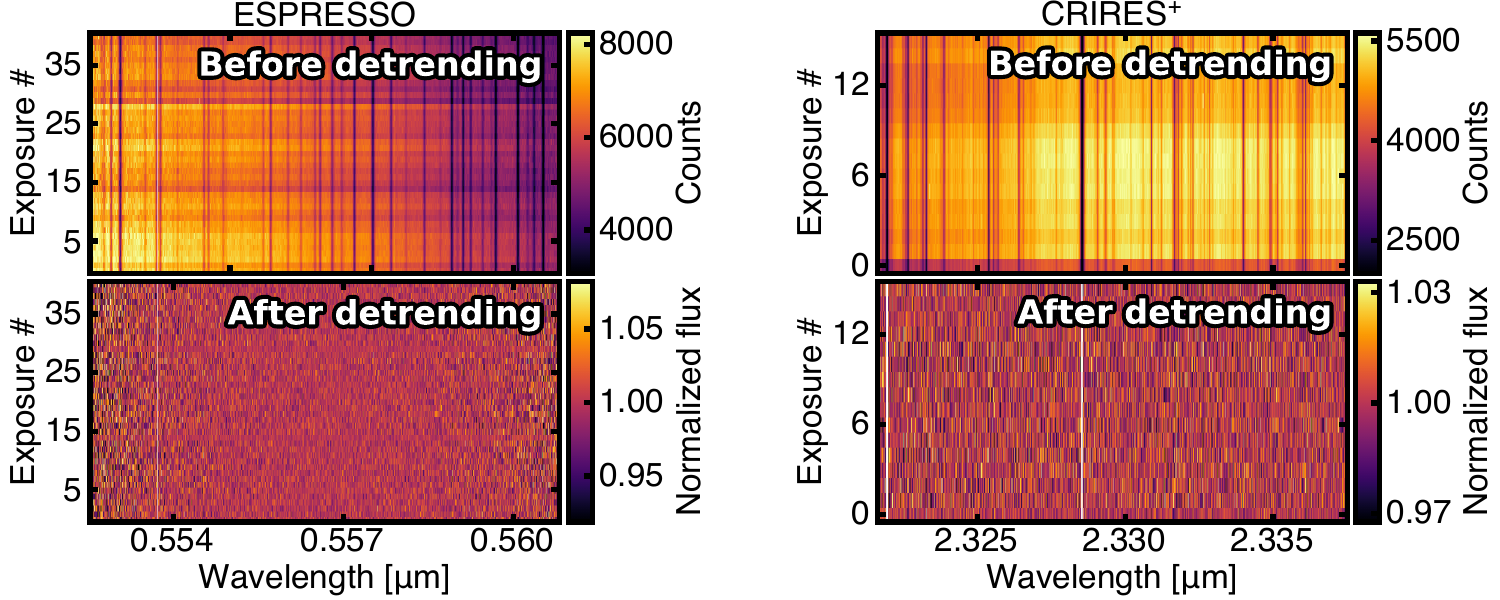}
\end{center}
\vspace{-4mm}
\caption{
Data detrending procedure for example spectral orders from the ESPRESSO (left) and CRIRES$^{+}$ (right) data sets.
The top left panel shows a segment of the extracted ESPRESSO spectral time series on 2021-01-11. Here darker vertical lines depict telluric and stellar absorption lines while horizontal stripes are due to variations in throughput, for example due to passing thin clouds.  The bottom left panel shows the data after the detrending steps (see Section~\ref{sec:methods}) are applied to remove unwanted flux contributions from everything other than WASP-121b.  The right panels shows the equivalent but for the CRIRES$^{+}$ time series of A nodding positions of 2022-02-13.  White regions denote wavelengths masked due to strong telluric absorption.  
The random noise appearance and lack of any distinguishable patterns in the data residuals indicates that the PCA (with three principal components removed in this case) efficiently removes both stellar and telluric contributions.  The bottom panels (`After detrending') are single-order examples of the data product used for the cross-correlation and retrieval analyses. 
}
\label{fig:data_detrending}
\end{figure*}
\begin{figure}[t]
\begin{center}
\includegraphics[width=\linewidth]{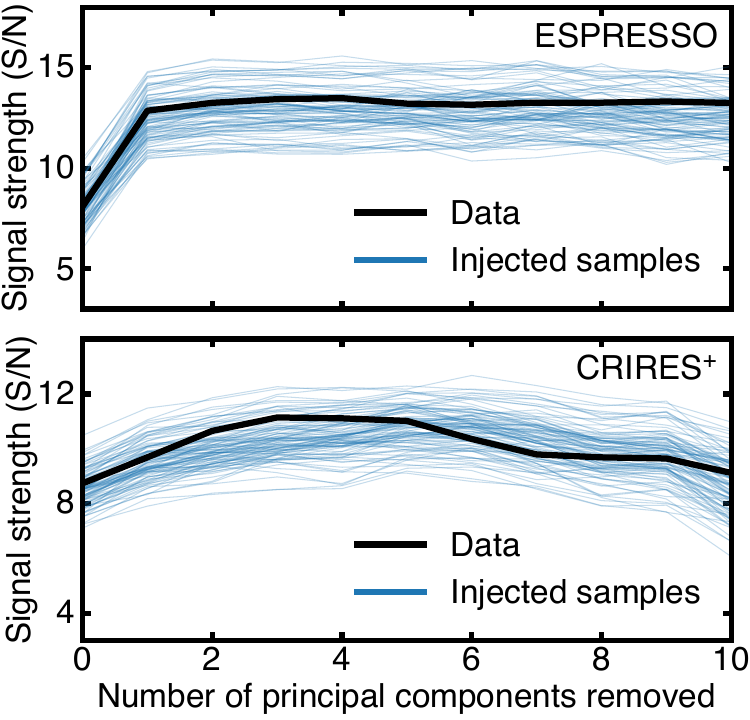}
\end{center}
\vspace{-4mm}
\caption{
Recovered cross-correlation S/N as a function of number of principal components removed in the data detrending procedure.  The black line shows the real signal from the atmosphere of WASP-121b observed with ESPRESSO (top) and CRIRES$^{+}$ (bottom), while the blue lines correspond to recovered models injected at random combinations of $K_p$ and $V_{\mathrm{sys}}$ away from the expected orbital position.  The atmosphere of WASP-121b is strongly detected irrespective of the number of components removed, and behaves similarly to injected models at different orbital configurations.  
Nevertheless, removing at least a few principal components to correct for telluric/stellar residuals significantly helps to reveal any underlying signal in the data.  However, removing too many can cause the PCA to `eat away' at the planetary signal and do more harm than good. For the main retrieval analysis in this work we remove three principal components from each data set, which best recovers both the real signals, and injected signals at surrounding orbital positions. 
}
\label{fig:optimal_PCA}
\end{figure}

Before attempting to detect the atmospheric signature from the atmosphere of WASP-121b, the data must be detrended to remove unwanted contributions from the tellurics and stellar lines that dominate the signal.  This is possible given that the line-of-sight orbital velocity of WASP-121b is rapidly changing relative to both the telescope rest frame and WASP-121 (the star).  Removing telluric and stellar lines is crucial as molecules such as H$_2$O are present in Earth's atmosphere, and metals such as Fe are present in the host star, and could otherwise contaminate any extracted signal.  The detrending procedure is done as follows:

\begin{enumerate}
    \item  For each night, the extracted spectra are organised in $N_{\mathrm{exp}} \times N_{\mathrm{order}} \times N_{\mathrm{pixel}}$ data cubes, with all ensuing operations done order-by-order.  An example time series for one epoch of each instrument is shown in Figure~\ref{fig:data_detrending} (top panels).
    \item  Outlier pixels in the spectroscopic time series deviating by more than 6$\sigma$ are flagged and corrected following \cite{brogi_carbon_2014}.  
    \item  Regions of heavy telluric contamination are masked out and not considered for the remainder of the analysis (Figure~\ref{fig:cross_sections}, top panels).  This includes the entirety of the 0.687--0.694\,$\mu$m, 0.76--0.77\,$\mu$m and 1.9--1.97\,$\mu$m wavelength ranges.  We further mask out wavelengths where the flux is less than 30\% of the continuum.  This latter masking process primarily affects the CRIRES$^{+}$ data due to tellurics in the infrared. 
    \item The ESPRESSO data (provided in the barycentric frame) is interpolated to the stellar rest frame, correcting for the gravitational reflex motion of WASP-121 caused by WASP-121b as it moves along its orbit.  This ensures that stellar lines are fixed in wavelength space for the detrending procedure.
    Meanwhile, the CRIRES$^{+}$ data is kept in the Earth's rest frame such that absorption from telluric lines always fall on the same pixels.  The choice of having the ESPRESSO and CRIRES$^{+}$ data sets in different rest frames is because stellar lines dominate the spectrum in the optical, while tellurics are the main source of contamination in the infrared.
    \item All spectra are continuum-aligned using a smoothing box $+$ Gaussian filter following \cite{gibson_relative_2022}.  This accounts for any throughput or blaze variations during time series observations and brings all exposures to the same continuum level. 
    \item As in \cite{brogi_retrieving_2019}, a second-order polynomial fit of the median spectrum is then divided out from each individual exposure. This acts as a first-order cleaning of the data but generally still leaves some contamination from stellar and telluric lines which are not perfectly constant-in-time.
    \item On top of the fitted out median spectrum, Principal Component Analysis (PCA) in the wavelength domain is used to remove any remaining unwanted residuals in the data.
    We tested removing different number of principal components (between 0 and 10) to ensure that any inferred signal is robust and independent of the exact number of components removed.  We find that removing at least 1--2 components is helpful to significantly reduce stellar and telluric residuals and better uncover the planetary signal.  However, applying PCA too aggressively can remove the underlying planet signal, especially for epochs with fewer exposures.  
    For our main analysis, we opt to remove three principal components from all time series', which we find nominally recovers injected signals at different orbital locations (Figure~\ref{fig:optimal_PCA}) in a methodology similar to what was done in \cite{holmberg_first_2022}. 
    \item Remaining noisy columns with a standard deviation greater than 4 times the median of their spectral order are masked and removed from the analysis.
    \item An uncertainty at every pixel to be used for the ensuing cross-correlation and retrieval analyses is estimated using a noise model as prescribed in \cite{gibson_detection_2020}.
\end{enumerate}

After these steps, each data time series should now be ridden of all unwanted pseudo-fixed-in-wavelength stellar and telluric contaminants (Figure~\ref{fig:data_detrending}, bottom panels), with any underlying rapidly accelerating atmospheric signal from WASP-121b still present ready to be uncovered and characterized.
Typically, the individual spectral features emitted from the atmosphere of WASP-121b are below the noise level too weak to see by eye, and can instead be revealed by stacking their contributions via a cross-correlation function~\citep[e.g.,][]{birkby_spectroscopic_2018}.

To avoid further interpolation of the data, we keep the ESPRESSO data in the stellar rest frame and the CRIRES$^{+}$ data in the telluric rest frame. 
We then re-align these in the cross-correlation analysis by interpolating the CCFs, or in the retrieval analysis by adjusting the Doppler shift applied to the model for each exposure.

\subsection{Classical cross-correlation analysis} \label{subsec:CCF}

With the cleaned CRIRES$^{+}$ and ESPRESSO data sets, we perform a classical cross-correlation analysis~\citep{snellen_orbital_2010, brogi_signature_2012, birkby_discovery_2017} to identify species present in the dayside atmosphere of WASP-121b.  
For this we first use SCARLET~\citep{benneke_atmospheric_2012, benneke_how_2013,  benneke_strict_2015, benneke_sub-neptune_2019, benneke_water_2019} to generate a thermal emission model of the dayside atmosphere WASP-121b, with a self-consistently generated inverted temperature structure assuming a solar composition in chemical equilibrium.  
Other than H$_2$-H$_2$, H$_2$-He collision-induced absorption, the model uses opacities for H$_2$O~\citep{polyansky_exomol_2018}, CO~\citep{rothman_hitemp_2010, li_rovibrational_2015}, OH~\citep{rothman_hitemp_2010}, SiO~\citep{yurchenko_exomol_2022}, CH$_4$~\citep{hargreaves_accurate_2020}, CO$_2$~\citep{yurchenko_exomol_2020}, HCN~\citep{barber_exomol_2014,harris_improved_2006}, C$_2$H$_2$~\citep{chubb_exomol_2020}, VO~\citep{mckemmish_exomol_2016}, TiO~\citep{mckemmish_exomol_2019}, as well as atomic metals Fe, Ni, Ca, Cr, Mn, and V~\citep{kurucz_including_2017}, and also H$^{-}$ (bound-free and free-free)~\citep{gray_observation_2021}.  Here the H$^{-}_{\mathrm{bf}}$ opacity depends on the abundance of hydride (H$^{-}$) while the H$^{-}_{\mathrm{ff}}$ opacity is a product of the abundance of neutral hydrogen atoms and the electron pressure set by the abundance of free electrons (e$^{-}$)~\citep{gray_observation_2021}.
Cross sections were computed using \texttt{HELIOS-K}~\citep{grimm_helios-k_2015, grimm_helios-k_2021} except for the case of atomic species contributed to \texttt{petitRADTRANS}~\citep{molliere_petitradtrans_2019} by K.\ Molaverdikhani\footnote{\url{https://petitradtrans.readthedocs.io/en/latest/content/available\_opacities.html}}, using the line lists and coefficients from R.\ Kurucz\footnote{\url{http://kurucz.harvard.edu/}} which include the effects of pressure broadening.  Chemical equilibrium abundances are calculated using \texttt{FastChem}~\citep{stock_fastchem_2018, stock_fastchem_2022}.  The fiducial model ($R$ = 250,000) is then convolved with first a rotational kernel assuming synchronous rotation with a limb velocity of 6.99\,km\,s$^{-1}$, followed by a Gaussian broadening kernel matching the instrumental resolutions ($R$ = 120,000 or $R$ = 105,000 for CRIRES$^{+}$ and $R$ = 140,000 for ESPRESSO).

\begin{figure*}
\begin{center}
\includegraphics[width=\linewidth]{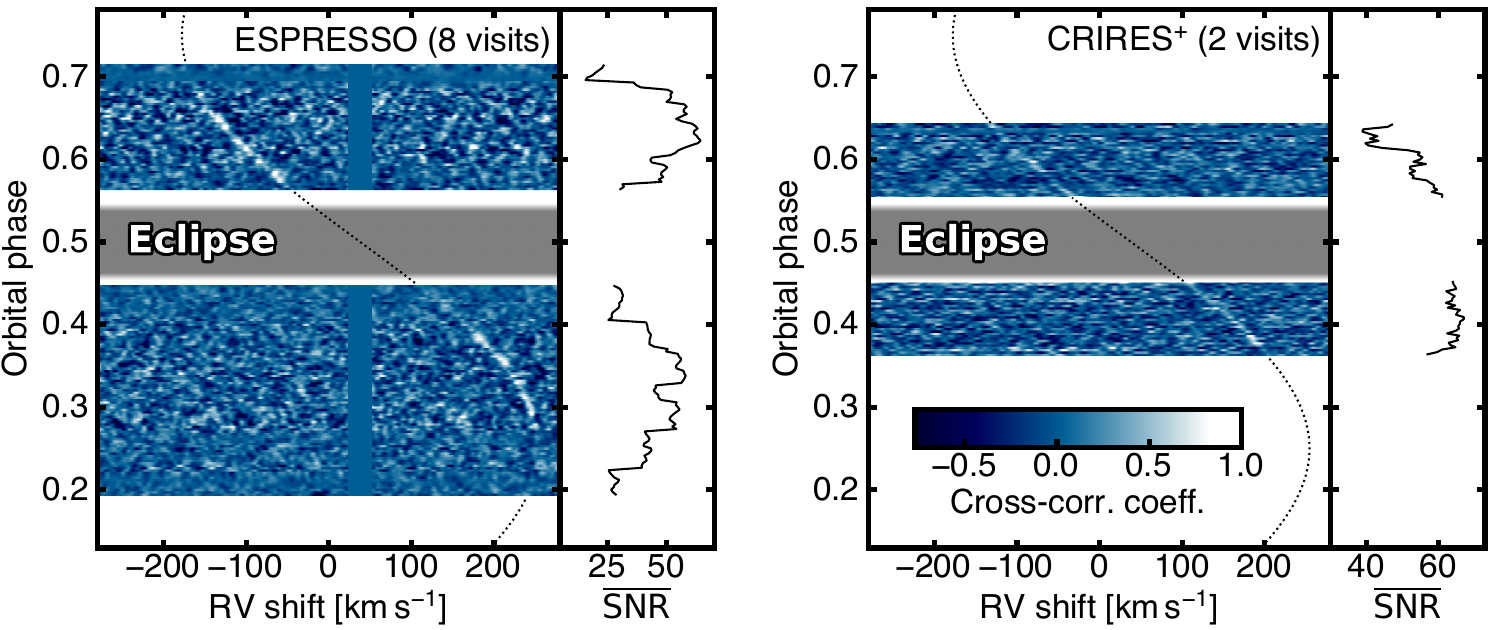}
\end{center}
\vspace{-4mm}
\caption{
Orbital signature of WASP-121b as seen from ESPRESSO and CRIRES$^{+}$.
The panels show the normalized cross-correlation as a function of phase for ESPRESSO (left) and CRIRES$^{+}$ (right). Here the signal from the thermal light of the dayside atmosphere of WASP-121b can be seen as a white streak closely following the expected orbital motion (dotted black line).  The shaded gray area marks the secondary eclipse, when data collection is avoided as WASP-121b passes behind its host star and is no longer visible.  
The ESPRESSO observations, comprising of eight visits totally 28 hours, cover a greater portion of the orbit than the 5.6 hours of observations split over two epochs using CRIRES$^{+}$.
The side panels denote the average signal-to-noise per wavelength at each observed phase.  
For ESPRESSO, the observed planetary trace naturally appears stronger at orbital phases of higher SNR that are observed more than once, and appears weaker at phases of lower SNR (covered by only a single visit) or close to quadrature when the planetary acceleration is minimal.  
To avoid any contamination of the signal, residuals near the systemic velocity (around 38\,km\,s$^{-1}$) due to the host star also having Fe in its atmosphere are masked. 
For CRIRES$^{+}$ the post-eclipse signal appears slightly weaker than the pre-eclipse signal, but this may be a results of the lower SNR of those data due to worse weather conditions.  
}
\label{fig:CCF_trail}
\end{figure*}

\begin{figure*}
\begin{center}
\includegraphics[width=\linewidth]{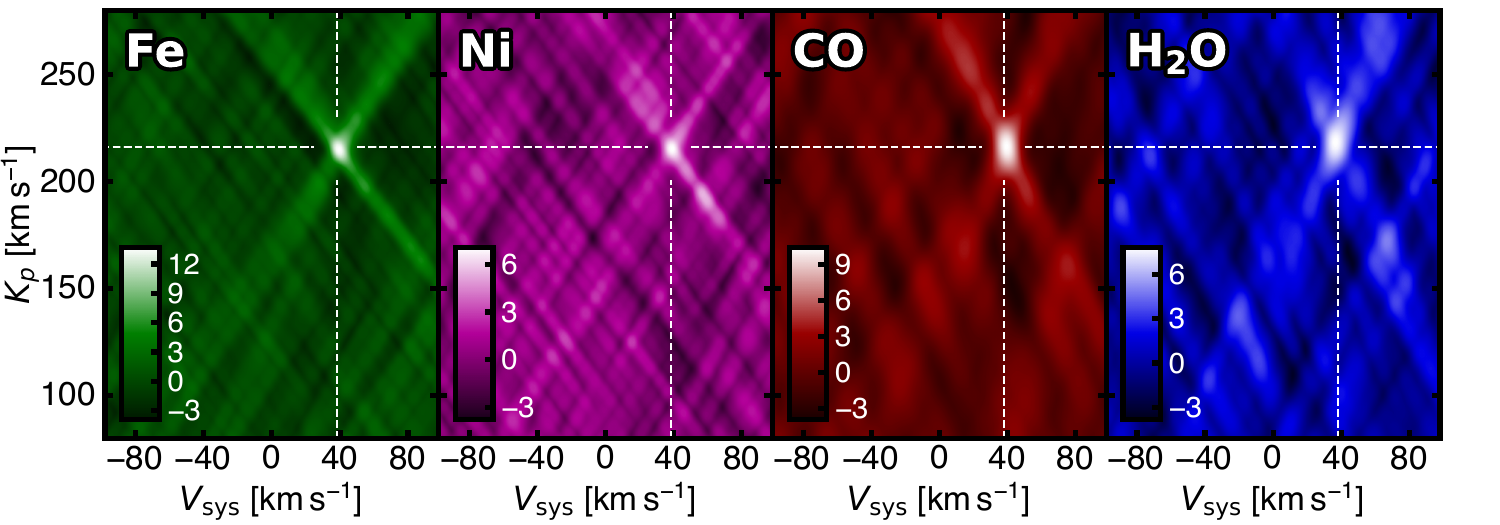}
\end{center}
\vspace{-4mm}
\caption{
Cross-correlation detections of Fe, Ni, CO, and H$_2$O in the dayside atmosphere of WASP-121b.
Here each panel depicts the two-dimensional cross-correlation signal-to-noise map of a species relative to the expected $K_p$ and $V_{\mathrm{sys}}$ (dashed white lines).  
Clear detections of Fe, Ni, CO, and H$_2$O can be seen as white blobs near the expected orbital and systemic velocities of WASP-121b. The atomic metal (Fe and Ni) detections are driven by the ESPRESSO data at optical wavelengths while the molecule (CO and H$_2$O) detections are primarily from the CRIRES$^{+}$ data in the near-infrared.  
The `X' shape of the detections is a result of observations having been obtained both pre- and post-eclipse, with the top left to bottom right diagonal being from the pre-eclipse data and the bottom left to top right diagonal being from the post-eclipse data.
While the Fe, Ni, and CO signals are well aligned with each other, H$_2$O is slightly blueshifted in comparison, potentially because of a combination of winds and a molecular dissociation-driven asymmetric distribution of water in the dayside atmosphere of WASP-121b. 
}
\label{fig:CCF_maps}
\end{figure*}

With an atmospheric template in hand, we compute the cross-correlation function (CCF) between the cleaned data and model contributions of individual species for each spectral time series. The CCFs are then summed together for all orders, interpolated on a common orbital phase grid, and then summed together for different epochs to produce combined `trail plots' showing the atmospheric signature of WASP-121b as a function of orbital phase (e.g., Figure~\ref{fig:CCF_trail}).  
Even though the relative radial velocity position of any potentially remaining residuals from metals in the stellar photosphere never overlaps with the planetary trace, to ensure that our results are not biased by these, we mask out CCF values within 15\,km\,s$^{-1}$ of the WASP-121b systemic velocity (38.2\,km\,s$^{-1}$, \cite{borsa_atmospheric_2021}).
The time-resolved cross-correlation functions are then phase folded to produce $K_p$-$V_{\mathrm{sys}}$ diagrams (e.g., Figure~\ref{fig:CCF_maps}), with signal-to-noise (S/N) estimates made by dividing the resulting cross-correlation maps by their standard deviation away from the signal of interest. 

A cross-correlation analysis provides a good first look at what species are present at a detectable level in an exoplanetary atmosphere, and whether dynamics or chemical/temperature/cloud inhomogeneities may be at play through deviations from the expected orbital velocities~\citep[e.g.,][]{ehrenreich_nightside_2020, wardenier_decomposing_2021, wardenier_modelling_2023, wardenier_phase-resolving_2024, kesseli_atomic_2022, pino_gaps_2022, prinoth_titanium_2022, savel_no_2022, vansluijs_carbon_2023, boucher_co_2023, pelletier_vanadium_2023, nortmann_crires_2024}. 
However, the classical cross-correlation approach provides only limited quantitative insights into the physical properties of the atmosphere of the planet of interest.  This is where high resolution atmospheric retrievals~\citep[e.g.,][]{brogi_framework_2017, brogi_retrieving_2019, gandhi_hydra-h_2019, gibson_detection_2020, pelletier_where_2021, line_solar_2021} can become powerful tools.

\subsection{Bayesian retrieval approach} \label{subsec:retrieval}

To further characterize the properties of the atmosphere of WASP-121b in a quantitative manner, we also perform retrieval analyses using the likelihood framework of \cite{gibson_detection_2020}.
One aspect that is unique to high-resolution atmospheric retrievals relative to standard analyses of space-based observations is that when processing ground-based high-resolution data to remove telluric or stellar features, it is possible to slightly alter the underlying planetary signal~\citep[e.g.,][]{brogi_retrieving_2019}.  In order to avoid any biases in any inferred parameters, it is therefore necessary to reproduce this effect on each sampled model so that it can represent the real underlying signal in the data as well as possible.  To do this, we apply the following steps for each likelihood calculation in the retrieval process.
\begin{enumerate}
    \item A SCARLET model of the thermal emission of WASP-121b is generated for a given set of parameters suggested by the retrieval. 
    \item The atmosphere model is convolved with kernels to account for both the rotation of the planet and the CRIRES$^{+}$ or ESPRESSO instrumental resolution.
    \item The model is then projected in time for the suggested $K_p$ and $V_{\mathrm{sys}}$ to account for the motion of the planet at the time stamp of each observed exposure.  
    As the optical and infrared data sets are in different rest frames (Section~\ref{subsec:detrending}), the Doppler shift applied to the model for each exposure also includes the stellar reflex motion in the case of the ESPRESSO data that is in the stellar rest frame and the barycentric Earth radial velocity (BERV) for the CRIRES$^{+}$ data that is in the telluric rest frame.
    Note that this and the following steps are unique for each of the ten observed time series'.
    \item Each modelled time series is multiplied by a phase function $(0.5(1 + \mathrm{cos}(2\pi\phi - \pi)))^2$ \citep[][see their Figure 3]{herman_dayside_2022, hoeijmakers_mantis_2024} to account for the expected variation in the observed brightness of WASP-121b depending on its orbital position $\phi$.
    \item A PHOENIX model~\citep{husser_new_2013} of the stellar spectrum ($T_{\mathrm{eff}}$ = 6500\,K, log$g$ = 4.0, Fe/H = 0.0) convolved to its projected rotational velocity $v\mathrm{sin}i = 13.5$\,km\,s$^{-1}$ \citep{delrez_wasp-121_2016}, Doppler shifted to the velocity of the WASP-121 system ($v_{\mathrm{sys}}$), and broadened to the instrumental resolution is used to scale the planetary emission at each time stamp to the expected planet-to-star flux contrast ratio.
    \item A box filter is further applied to account for the motion of WASP-121b within the span of each exposure.  This added `exposure blurring' is particularly important for longer exposures, short orbital period exoplanets, and for observations when the planetary acceleration from our line-of-sight is maximal (near transit and secondary eclipse).  
    In this case, WASP-121b can move by up to $\sim$3\,km\,s$^{-1}$ within a single (300s) exposure near superior conjunction.
    \item The generated exposure-dependent model is then injected in a reconstruction of everything that was removed from getting from the top panel to the bottom panel of Figure~\ref{fig:data_detrending}, for each night.
    \item Steps 6 and 7 of the data detrending procedure (Section~\ref{subsec:detrending}) are repeated on each reconstructed time series to reproduce the effects of the detrending on the model.
    \item The log likelihood between the cleaned data and the now modified model time series is computed and summed for all observations.  As no continuum information is retained from the normalized ground-based high-resolution observations~\citep{brogi_retrieving_2019}, contributions from different spectral orders or instruments can naturally be combined together. This is in contrast to space-based observations that retain absolute flux calibration but can require offsets to be fitted to combine data sets from different instruments~\citep[e.g.,][]{benneke_sub-neptune_2019, moran_high_2023, madhusudhan_carbon-bearing_2023, welbanks_high_2024}.

\end{enumerate}

Similar to previous works using this framework~\citep{pelletier_where_2021, pelletier_vanadium_2023, bazinet_subsolar_2024}, we fit for the atmospheric composition, vertical temperature structure, and orbital parameters of WASP-121b simultaneously.  For the temperature-pressure profile, we use the free parameterization from \cite{pelletier_where_2021}, fitting 10 temperature points uniformly distributed in log pressure between 1 and 10$^{-8}$\,bar with a smoothing prior $\sigma_{\mathrm{smooth}}$ = 300\,K\,dex$^{-2}$.  We also fit for the orbital parameters $K_p$ and $V_{\mathrm{sys}}$.

To determine the atmospheric composition of WASP-121b, we perform two types of retrievals. (1) A hybrid free retrieval similar to \cite{coulombe_broadband_2023} where each species is fit individually and abundances are assumed to be constant-with-altitude except for H$_2$O and H$^{-}$ for which the abundance profiles are parameterized using a power law following \cite{parmentier_thermal_2018} (see their Table 1) to take into account the strong dependence of H$_2$O and H$^{-}$ on pressure and temperature in ultra-hot Jupiter atmospheres. 
We adopt such a modification because, similar to \cite{gandhi_revealing_2024, mansfield_metallicity_2024}, we find that a traditional free retrieval fitting a uniform H$_2$O profile without taking into account the thermal dissociation of water molecules infers extremely low H$_2$O abundances that likely do not represent the composition of the deeper atmosphere.
(2) A retrieval enforcing chemical equilibrium, varying the C/O ratio and separate metallicities for volatile species ($[$M$_{\mathrm{vol}}$/H$]_{\odot}$), refractory species ($[$M$_{\mathrm{ref}}$/H$]_{\odot}$), and metal oxides TiO and VO ($[$M$_{\mathrm{TiO, VO}}$/H$]_{\odot}$).  Here the metallicities are with respect to solar and on a log scale, so for example $[$M$_{\mathrm{ref}}$/H$]_{\odot}$ = 0.5 would correspond to a refractory metallicity of 10$^{0.5}\times$ solar.
The choice of fitting a separate metallicity for TiO and VO is because these species have both not yet been robustly detected in the atmosphere of WASP-121b, and are known to have imperfect line lists~\citep[e.g.,][]{merritt_non-detection_2020, regt_quantitative_2022, maguire_high_2024}.  TiO has also been suggested to be cold-trapped on the nightside of WASP-121b~\citep{hoeijmakers_hot_2020, hoeijmakers_mantis_2024}, and hence could otherwise bias the retrieved refractory metallicity if included in $[$M$_{\mathrm{ref}}$/H$]_{\odot}$ but not actually present in the gas phase on the dayside.  However, we still allow for the retrieval to include TiO and VO as these have strong opacity bands in the optical and could potentially be important pseudo-continuum contributors to the underlying planetary spectrum.
While not as flexible as a free retrieval, a chemical equilibrium retrieval accounts for the thermal dissociation of molecules and the ionization of metals
in a chemically consistent manner.


In the hybrid free retrieval we fit for the abundances of the primary carriers of carbon, oxygen, iron, and nickel expected for ultra-hot Jupiters that these data are sensitive to (CO, H$_2$O, Fe, Ni), as well as the main expected sources of continuum opacity (H$^{-}_{\mathrm{bf}}$ and H$^{-}_{\mathrm{ff}}$).  The hydride (H$^{-}$) abundance fitted in the retrieval controls the bound-free continuum important at optical wavelengths while the abundance of free electrons (e$^{-}$), together with the neutral atomic hydrogen (H) abundance, sets the free-free continuum opacity strongest at longer wavelengths (Figure~\ref{fig:cross_sections}, bottom panel). We note that while the free retrieval has the flexibility advantage of not being restricted by prior assumptions on the chemistry, ionization, and relative abundance of different metals, the number of included species, each as a free parameter, can quickly run into computational limitations.
However, we do include opacity contributions from additional molecules and metals in the chemically consistent retrievals.

For the chemical equilibrium retrieval, we include additional opacity contributions from OH, HCN, C$_2$H$_2$, CH$_4$, CO$_2$, SiO, Cr, Mg, V, Ca, Mn, VO, and TiO to the atmosphere model.  However, for a given queried temperature structure and C/O ratio, abundance profiles are predicted with \texttt{FastChem} assuming different metallicities for different species.  Specifically, volatile molecules H$_2$O, CO, OH, HCN, C$_2$H$_2$, CH$_4$, CO$_2$ are controlled by the volatile metallicity $[$M$_{\mathrm{vol}}$/H$]_{\odot}$, refractory-bearing species Fe, Ni, Cr, Mg, V, Ca, Mn, and SiO follow the refractory metallicity $[$M$_{\mathrm{ref}}$/H$]_{\odot}$, while the abundance of VO and TiO are calculated assuming a separate metallicity $[$M$_{\mathrm{TiO, VO}}$/H$]_{\odot}$.  Meanwhile, all other elemental abundance ratios are assumed to be in solar proportion~\citep{asplund_chemical_2009} relative to oxygen, expect for carbon which is varied via the fitted C/O ratio.

\begin{table}
\caption{Atmospheric retrieval parameter description and prior ranges for both the hybrid free and chemical equilibrium retrievals.} 
\label{tab:retrieval_priors}
\centering
\def\arraystretch{1.1}
\begin{tabular}{ccc}
\hline
\hline
Parameter & Description & Prior range \\
\hline
\hline
$K_p  $ & Keplerian velocity [km\,s$^{-1}$]  & 200 to 240 \\
$V_\mathrm{sys}  $ & Systemic velocity [km\,s$^{-1}$]  & $30$ to $50$ \\
$T_i  $ & Temperature of layer $i$ [K] & $100$ to $8500$ \\
\hline
 & Hybrid free & \\
\hline
$\log \chi_j$ & VMR of species $j$ & $-12$ to $0$ \\
\hline
 & Chemical equilibrium & \\
\hline
$[$M$_{\mathrm{vol}}$/H$]_{\odot}$ & Volatile metallicity & $-3$ to $3$ \\
$[$M$_{\mathrm{ref}}$/H$]_{\odot}$ & Refractory metallicity & $-3$ to $3$ \\
$[$M$_{\mathrm{TiO, VO}}$/H$]_{\odot}$ & TiO + VO metallicity & $-3$ to $3$ \\
C$/$O & Carbon-to-oxygen ratio & $0$ to $10$ \\
$\log P_\mathrm{c}$ & Continuum pressure [log\,bar] & $-8$ to $2$ \\
\hline
\multicolumn{3}{l}{\small $i$ = 0 to 9 ($T_{0}$ at $10^{-8}$\,bar and $T_9$ at $10$\,bar)}\\
\multicolumn{3}{l}{\small $j$ = H$_2$O, CO, Fe, Ni, H$^{-}$, e$^{-}$}
\end{tabular}\\ 
\end{table}

The chemically consistent retrieval also naturally takes into account atomic C and O when computing the carbon and oxygen budgets. This is particularly important when probing the lower pressure regions of thermally inverted ultra-hot Jupiter atmospheres~\citep[e.g.,][]{brogi_roasting_2023}. WASP-121b is also known to have Fe$^{+}$ in its upper atmosphere, as seen from transit observations~\citep{maguire_high-resolution_2023}.
Although ionization of Fe into Fe$^{+}$ is taken into account in our chemical equilibrium retrieval, we find that this does significantly affect our abundance inference as our thermal emission data primarily probe regions of the atmosphere where Fe is still predominantly in neutral form (Figure~\ref{fig:TP_VMRs}, top right panel).

Due to its elevated ($>$2500\,K) dayside temperature~\citep{bourrier_optical_2020, mikal-evans_diurnal_2022, mikal-evans_jwst_2023, changeat_is_2024}, WASP-121b is expected to be too hot to allow for the formation of clouds~\citep{parmentier_exoplanet_2018, gao_aerosol_2020, helling_cloud_2021} on its dayside. 
We therefore fit for H$^{-}$ as a source of continuum opacity (both bound-free and free-free) in the retrievals, which is more applicable for ultra-hot environments~\citep{arcangeli_h-_2018, gandhi_h-_2020}. 
However, given that clouds may still be present on the colder nightside, and a possible explanation for the apparent lack of any ultra-refractory species that have very high condensation temperatures (such as titanium) yet robustly detected on WASP-121b~\citep{merritt_non-detection_2020, merritt_inventory_2021, hoeijmakers_hot_2020, hoeijmakers_mantis_2024, maguire_high-resolution_2023, gandhi_retrieval_2023}, we also test fitting for a gray continuum opacity pressure $P_c$ in the retrieval. Practically, for these emission spectra, $P_c$ can act as a pressure level below which either the temperature structure becomes isothermal or a gray cloud deck may exist.

In ultra-hot Jupiters, molecular hydrogen is also expected to thermally dissociate into atomic hydrogen in hotter and higher altitude regions of the atmosphere, which can cause a significant change in the local atmospheric scale height~\citep{bell_increased_2018, komacek_effects_2018, parmentier_thermal_2018, wardenier_decomposing_2021}.  While this is naturally computed in the chemical equilibrium retrieval, we account for this in our hybrid free retrieval as in \cite{pelletier_vanadium_2023} by using the proportions of H, H$_2$, and He predicted by a \texttt{FastChem} computation made for each queried temperature structure as filler gases in the atmosphere model (such that the volume mixing ratio of all gases sum to one in all atmospheric layers).  In general, this results in the atmosphere being H-dominated in the pressure regions probed, with H$_2$ molecules being more present in deeper layers.  The predicted H abundance, along with the calculated electron pressure, is also used to compute the H-e$^{-}$ free-free continuum~\citep{gray_observation_2021}.

The Bayesian inference and parameter exploration is done using the \texttt{emcee} Markov chain Monte Carlo Python package~\citep{foreman-mackey_emcee_2013}. Priors of the fitted parameters are assumed to be flat, either in normal space for the velocity parameters, C/O ratio, and temperature points, or in log space for the different metallicities, volume mixing ratios of individual species, and the continuum pressure level (Table~\ref{tab:retrieval_priors}).


\section{Results and discussion}\label{sec:results}

\subsection{Cross-correlation detections} \label{subsec:ccf_results}

We strongly detect the phase-resolved orbital trace of the dayside atmosphere of WASP-121b with both ESPRESSO and CRIRES$^{+}$ (Figure~\ref{fig:CCF_trail}).  All signals are observed in emission, indicative of a thermally inverted atmosphere, consistent with previous works~\citep{evans_ultrahot_2017, mikal-evans_confirmation_2020, mikal-evans_diurnal_2022, hoeijmakers_mantis_2024, changeat_is_2024}.
Searching for individual species, we detect CO, H$_2$O, Fe, and Ni which all fall close to the expected orbital location of WASP-121b (Figure~\ref{fig:CCF_maps}).
The detections of volatile molecules CO and H$_2$O originate from the CRIRES$^{+}$ data in the near-infrared.
Meanwhile, the signals from refractory atomic Fe and Ni are driven by the ESPRESSO optical data, and are a confirmation of the detection of these species presented in \cite{hoeijmakers_mantis_2024}.   

Cross-correlation maps appear slightly more blurry for CO and H$_2$O compared to Fe and Ni primarily because CRIRES$^{+}$ has a slightly lower spectral resolving power than ESPRESSO.
Beyond the natural broadening due to dynamics and the planetary rotation, spectral lines originating from the atmosphere of WASP-121b also shift by up to several km\,s$^{-1}$ over the course of each exposure due to its rapid orbital motion, causing an additional velocity blurring effect.

While \cite{hoeijmakers_mantis_2024} also present a plethora of other detections on WASP-121b, we focus primarily on Fe and Ni in this work as these are the two refractory elements most strongly detected.  The motivation behind only considering Fe and Ni instead of the full inventory of atomic species detected on WASP-121b is because rock-forming elements are expected to behave similarly (i.e., always being in a solid state) under the protoplanetary disk conditions in which giant planets are expected to form~\citep{chachan_breaking_2023}.  Hence, in theory, Fe (or any other refractory metal) should be a representative tracer of the accreted refractory content of WASP-121b.  
However, we also include Ni in order to verify that we obtain consistent results regardless of which tracer of refractories is used.
We note that while molecules such as SiO and OH are likely present in the atmosphere of WASP-121b~\citep{lothringer_uv_2022, wardenier_phase-resolving_2024}, our observations cannot confirm or rule out their presence via a cross-correlation analysis as these species only have feeble spectral features in the CRIRES$^{+}$ and ESPRESSO wavelength ranges (Figure~\ref{fig:cross_sections}).

\subsection{Velocity offsets} \label{subsec:vel_offsets}

In velocity space, we find that while CO, Fe, and Ni are well aligned with the previously measured $V_{\mathrm{sys}}$ ($38.198\pm0.002$\,km\,s$^{-1}$) and $K_p$ ($216.24\pm0.51$\,km\,s$^{-1}$) values~\citep{borsa_atmospheric_2021, hoeijmakers_mantis_2024}, the H$_2$O signal is comparatively blueshifted (Figure~\ref{fig:CCF_maps}). We note that a slight ($<2$\,km\,s$^{-1}$) blueshift in H$_2$O emission relative to CO at dayside phases can naturally be explained by three dimensional effects and the inclusion of magnetic fields in simulations of the similar ultra-hot Jupiter WASP-76b~\citep{beltz_magnetic_2022}.  However, this is a bit less that the 2--3\,km\,s$^{-1}$ relative blueshift we observe on WASP-121b. Our observed H$_2$O signal is also positioned at a slightly higher $K_p$ value than the other species. Curiously, our measured relative H$_2$O blueshift is the opposite of what was found on the dayside of another ultra-hot Jupiter, WASP-18b, using IGRINS~\citep{brogi_roasting_2023}.

One possible explanation that could explain the offset of the water detection relative to the other species could be if the H$_2$O abundance is non-uniform on the dayside of WASP-121b, for example, if water dissociation occurs most severely at (and downwind of) the hottest regions of the atmosphere slightly eastwards of the substellar point~\citep{mikal-evans_jwst_2023}.  In this scenario CO, Fe, and Ni emit from all facets of the atmosphere while H$_2$O predominantly originates from cooler regions of the atmosphere westwards of the substellar point that appear blueshifted from our line-of-sight due to rotation and an eastward equatorial jet~\citep{showman_three-dimensional_2015}.

\begin{table}
\caption{Atmospheric retrieval results.} 
\label{tab:retrieval_results}
\centering
\def\arraystretch{1.1}
\begin{tabular}{ccc}
\hline
\hline
Parameter & Hybrid free & Chemical equilibrium~ \\
\hline
\hline
$K_p$ &  $216.51_{-0.33}^{+0.28}$  & $216.49_{-0.30}^{+0.29}$ \\
$V_\mathrm{sys}  $ & $38.35_{-0.23}^{+0.22}$ & $38.31_{-0.22}^{+0.21}$ \\
$\log \chi_{\mathrm{CO}}$ & $-2.62_{-0.27}^{+0.28}$ & -- \\
$\log \chi_{\mathrm{H}_2\mathrm{O}}$ & $-3.43_{-0.27}^{+0.26}$ & -- \\
$\log \chi_{\mathrm{Fe}}$ & $-4.07_{-0.26}^{+0.25}$ & -- \\
$\log \chi_{\mathrm{Ni}}$ & $-5.16_{-0.27}^{+0.24}$ & -- \\
$\log \chi_{\mathrm{H}^{-}}$ & $<-$5.85 &  -- \\
$\log \chi_{\mathrm{e}^{-}}$ &  $<-$1.02 &  -- \\
$[$M$_{\mathrm{vol}}$/H$]_{\odot}$ & $0.76_{-0.26}^{+0.28}$ & $0.64_{-0.11}^{+0.12}$ \\
$[$M$_{\mathrm{ref}}$/H$]_{\odot}$ & $0.41_{-0.25}^{+0.24}$ & $0.35_{-0.09}^{+0.10}$ \\
$[$M$_{\mathrm{TiO, VO}}$/H$]_{\odot}$ & -- & $<-$0.71 (3$\sigma$) \\
C$/$O & $0.87_{-0.06}^{+0.04}$ & $0.73_{-0.08}^{+0.07}$ \\
$\log P_\mathrm{c}$ & -- & $>-$1.55 (3$\sigma$) \\
\hline
$[$M$_{\mathrm{vol}}$/M$_{\mathrm{ref}}]_{\odot}$ & $0.35_{-0.14}^{+0.13}$ & $0.29\pm0.12$ \\
C/Fe & $29.1_{-8.0}^{+10.9}$ & $21.3_{-5.2}^{+7.2}$ \\
O/Fe & $34.0_{-9.8}^{+12.8}$ & $29.6_{-7.0}^{+9.6}$ \\
Ni/Fe & $0.08\pm0.02$ & -- \\
\hline
\end{tabular}\\ 
\end{table}

\begin{figure*}[t!]
\begin{center}
\includegraphics[width=\linewidth]{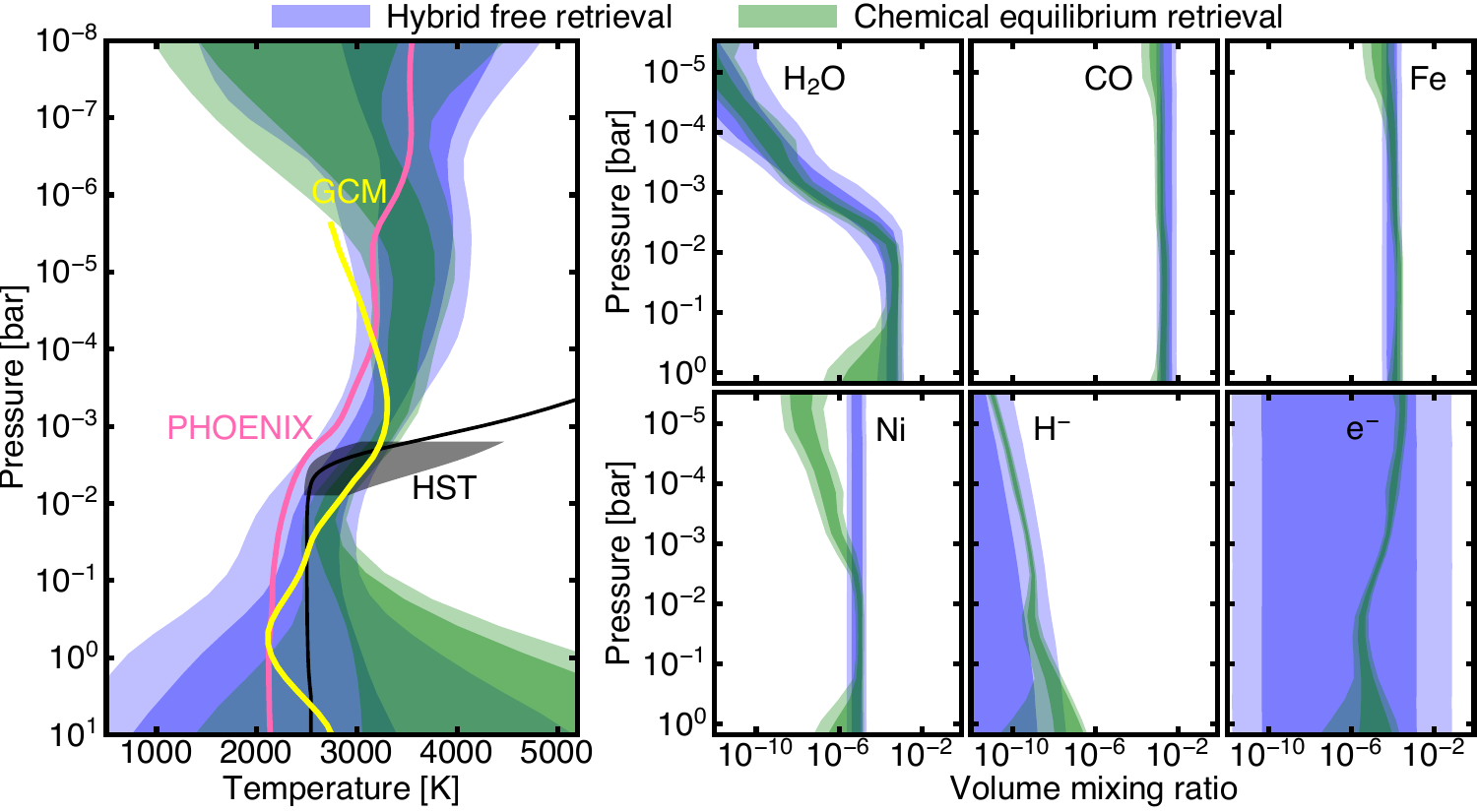}
\end{center}
\vspace{-3mm}
\caption{
Retrieved temperature structure and abundance profiles on the dayside atmosphere of WASP-121b.  The results from the hybrid free retrieval considering parameterized H$_2$O dissociation are shown in blue while the chemical equilibrium retrieval results are shown in green.
Left panel: inferred vertical temperature-pressure profile of WASP-121b (68\% and 95\% confidence interval contours) compared to a one-dimensional radiative-convective equilibrium PHOENIX model~\citep{lothringer_uv_2020} (pink solid line), a three-dimensional GCM~\citep{parmentier_thermal_2018} (yellow solid line), and previous results using the Hubble Space Telescope~\citep{mikal-evans_diurnal_2022} (black contour, 1$\sigma$ credible range at the pressures probed by the data).  
Right panels: Retrieved abundance profiles for H$_2$O, CO, Fe, Ni, H$^{-}$, and e$^{-}$ (in order from top left to bottom right).  In the hybrid free retrieval, abundance profiles are assumed to be uniform-in-pressure except for H$_2$O and H$^{-}$ which are fitted using the parameterization of \cite{parmentier_thermal_2018} to take into account thermal dissociation.  Here we see that, for H$_2$O in particular, dissociation is important at pressures below $\sim$10\,millibar.
Meanwhile, we find that for our retrieved temperature structure of WASP-121b, the chemical equilibrium abundance profile predicted for CO and Fe can still be reasonably well approximated as being constant-with-altitude.  This is in contrast to Ni, which ionizes more readily compared to Fe and for these temperatures is expected to appear underabundant at sub-millibar pressures.
With its greater flexibility, uncertainties are generally larger for the hybrid free retrieval, in particular for parameters for which we have little sensitivity to, such as the abundance of free electron.
Overall we find that the dayside atmosphere of WASP-121b has a strong thermal inversion and that our abundance estimates from both retrieval approaches are compatible.
}
\label{fig:TP_VMRs}
\end{figure*}

\begin{figure*}[t!]
\begin{center}
\includegraphics[width=\linewidth]{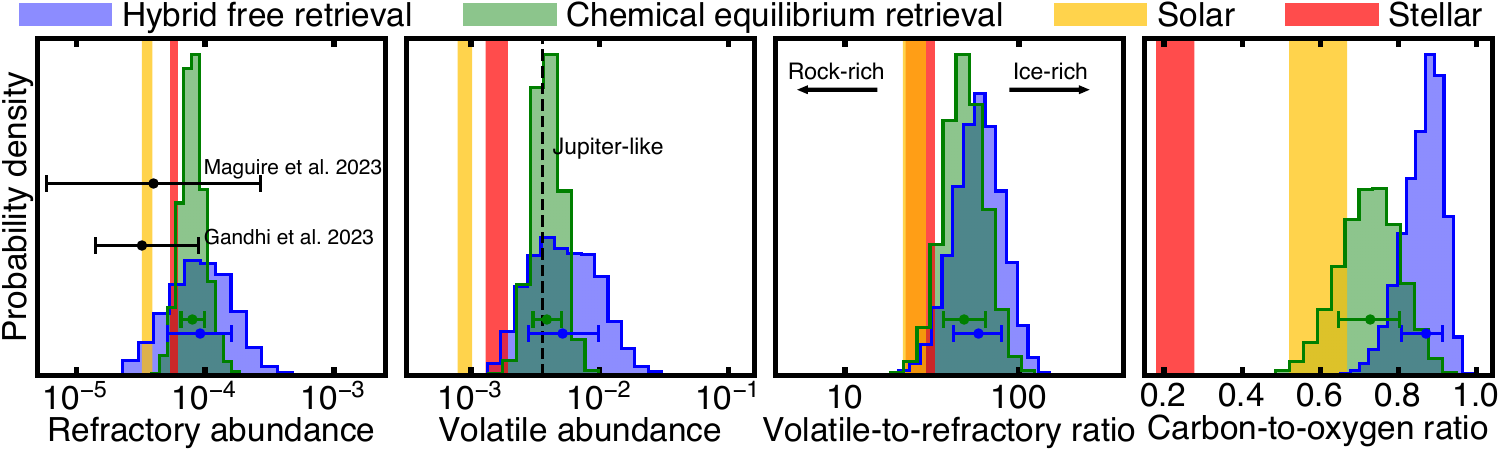}
\end{center}
\vspace{-3mm}
\caption{
Retrieved elemental composition of the dayside atmosphere of WASP-121b from both the free (blue) and chemical equilibrium (green) retrievals. 
Shown in the panel from left to right are the marginalized probability constraints for the inferred abundance of refractories, abundance of volatiles,  ratio of volatiles to refractories, and C/O ratio in the atmosphere of WASP-121b.  Values for the Sun (orange) and host star (red) are also depicted for comparison.  
While the abundance of refractories on WASP-121b is marginally elevated but consistent with previous works~\citep{gandhi_retrieval_2023, maguire_high-resolution_2023} and the stellar value to about one sigma, the abundance of volatiles is slightly super-stellar, more consistent with a Jupiter-like degree of enrichment ($\sim$4$\times$ solar, black dashed line).
Combined, the volatile-to-refractory ratio is elevated with respect to both solar and stellar values, suggesting that WASP-121b has an atmosphere that is enriched in ices relative to rocks.  
The retrieved high C/O ratio, especially in comparison with the host star, also potentially indicates that WASP-121b may have accreted its envelope in a carbon-rich environment.
Overall, both the free and chemical equilibrium retrievals find relatively consistent results, with the volatile and refractory abundances as well as the volatile-to-refractory and carbon-to-oxygen ratios on WASP-121b all being enriched relative to both the Sun and WASP-121 (the host star).
}
\label{fig:elemental_ratios}
\end{figure*}

\subsection{Retrieval results} \label{subsec:TP}

We find that WASP-121b has a pronounced stratosphere exceeding 3000\,K at pressures below one millibar (Figure~\ref{fig:TP_VMRs}, left panel).  The retrieved temperature structure is similar for both the hybrid-free and chemical equilibrium retrievals, 
mostly only showing larger differences at either very low or high pressures that are not well probed by the data.  

We compare our retrieved temperature-pressure profile to 
a GCM of WASP-121b generated using the SPARC/MITgcm framework~\citep{showman_atmospheric_2009, parmentier_3d_2013, parmentier_transitions_2016, parmentier_thermal_2018} assuming no clouds and a solar-like composition under equilibrium chemistry. 
The GCM broadly resembles the inferred temperature structure, with differences likely due to the assumed solar-like composition. We note that increasing the metallicity in the model would result in a shift in the lapse rate to lower pressures and drive a stronger thermal inversion (e.g., see Figure~5 of \cite{parmentier_thermal_2018}).  

We also verify how our retrieved temperature profile compares to a self-consistent one-dimensional radiative-convective-thermochemical equilibrium (1D-RCTE) model of the vertical temperature structure using the PHOENIX framework~\citep{barman_irradiated_2001}.  As in previous applications to ultra-hot Jupiters~\citep{lothringer_extremely_2018, lothringer_influence_2019, lothringer_phoenix_2020, lothringer_uv_2020, lothringer_uv_2022}, the PHOENIX model assumes full heat redistribution, an internal temperature of 200\,K, a solar-like composition, and includes opacity contributions from 130 chemical species, including molecules as well as neutral and ionized atoms.  
Compared to the GCM, the PHOENIX model contains opacity sources from many additional optical absorbers, enabling more heat to be absorbed in the upper atmosphere.  The overall higher temperatures retrieved at the pressure ranges probed by our data ($\sim$$10^{-5}$--$10^{-2}$\,bar) may indicate that heat redistribution on WASP-121b is not perfectly efficient as assumed in the model, consistent with previous findings from phase curve observations~\citep{bourrier_optical_2020, daylan_tess_2021, mikal-evans_jwst_2023}.

We stress that these forward models are not fitted to the data, but rather physically driven predictions meant to provide a first-order comparison with self-consistent atmospheric scenarios. The inferred temperature-pressure profile from the high-resolution ESPRESSO and CRIRES$^{+}$ data is also broadly consistent with previous results from HST WFC3~\citep{mikal-evans_diurnal_2022, changeat_is_2024}.

Our retrieved abundance profiles suggest that significant H$_2$O dissociation is ongoing on the dayside atmosphere of WASP-121b (Figure~\ref{fig:TP_VMRs}, top middle panel).  The more stable species CO and Fe are relatively well approximated as being constant-in-pressure even up to sub-millibar pressures, while Ni is likely partially ionized at these lower pressures. For the continuum contributors, only an upper limit is obtained for the H$^{-}$ abundance while e$^{-}$ is poorly constrained by these data.
Even though allowed to explore compositions forbidden by thermochemical equilibrium, abundance profiles inferred from the hybrid free retrieval are overall well consistent with those from the chemical equilibrium retrieval (Figure~\ref{fig:TP_VMRs}, right panels). The convergence to similar results using both retrieval approaches 
shows that both these methods, and their respective advantages, can be leveraged for characterizing ultra-hot Jupiter atmospheres.

We note that while absolute abundance parameters in the retrievals are strongly correlated to both the temperature structure and continuum opacity sources, relative abundance constraints from high-resolution spectroscopy are generally more robust~\citep{gandhi_retrieval_2023} (e.g., see their Section 3.3 for a detailed discussion).

From our chemical equilibrium retrieval, we also find that the metallicity controlling the abundances of TiO and VO is significantly depleted compared to the metallicity of other refractories (3$\sigma$ upper limit $[$M$_{\mathrm{TiO, VO}}$/H$]_{\odot}$ $<-$0.71 compared to $[$M$_{\mathrm{ref}}$/H$]_{\odot} = 0.41_{-0.25}^{+0.24}$).  This supports the idea that titanium may be cold-trapped on the colder nightside of WASP-121b~\citep{hoeijmakers_mantis_2024}.  We also rule out the presence of a gray continuum absorber at pressures below about 30 millibar at the 3$\sigma$ level ($\log P_\mathrm{c}$ $>-$1.55\,bar).

\subsection{Inferred elemental abundance ratios} \label{subsec:abundances}

We derive elemental abundance ratios for the atmosphere of WASP-121b, finding that the refractory abundance matches well a star-like~\citep{polanski_chemical_2022}, slightly super-solar~\citep{asplund_chemical_2021} composition (Figure~\ref{fig:elemental_ratios}, first panel).
Meanwhile, the volatile abundance is slightly more enhanced relative to both the Sun and host star, more comparable to a Jupiter-like ($\sim$4$\times$solar) level of enrichment~\citep{atreya_composition_2003} (Figure~\ref{fig:elemental_ratios}, second panel). 
In the hybrid free retrieval the volatile metallicity is determined from H$_2$O and CO while the refractory metallicity is inferred from Fe and Ni.
The chemistry retrieval further includes contributions from additional species and takes into account the ionization and dissociation (Figure~\ref{fig:TP_VMRs}).  
Combined, the volatile-to-refractory ratio, which is a proxy of the ice-to-rock ratio, is measured to be $1.75_{-0.41}^{+0.57}$ (chemical equilibrium retrieval) or $2.12_{-0.59}^{+0.77}$ (free retrieval) times more elevated than the stellar photosphere (Figure~\ref{fig:elemental_ratios}, third panel).
We also find that the atmosphere of WASP-121b is carbon-enriched, with a C/O ($0.87_{-0.06}^{+0.04}$ from the free retrieval and $0.73_{-0.08}^{+0.07}$ from the chemically consistent retrieval) that is both super-solar and super-stellar (Figure~\ref{fig:elemental_ratios}, right panel).

The abundance inferences are similar for the hybrid free retrieval parameterizing H$_2$O dissociation and the equilibrium chemistry retrieval, with both finding that WASP-121b is enriched in volatiles relative to refractories and an elevated C/O ratio. 
Although in the chemical equilibrium retrieval an overall refractory metallicity fits for the abundances of all refractory species together assuming solar proportions~\citep{asplund_chemical_2009}, the Fe and Ni abundances are fit independently in the free retrieval where we find a Ni/Fe ratio that is consistent with both that of the Sun and the host star to within about 2$\sigma$ ($\log$(Ni/Fe) = $0.17_{-0.11}^{+0.10}$ relative to solar and $0.24_{-0.12}^{+0.11}$ relative to stellar).
The consistency between Fe and Ni reinforces the idea that refractory elements should generally maintain near-stellar proportions in ultra-hot Jupiter atmospheres given that they are expected to all behave similarly in the protoplanetary disc~\citep{chachan_breaking_2023}.
Similarly, previous observations in transmission have shown that abundance ratios of Fe, Mg, Cr, and V on WASP-121b are consistent with a stellar-like composition~\citep{maguire_high-resolution_2023}. In general this means that the measurement of any one rock-forming element should act as an accurate proxy of the overall refractory metallicity. 
However, we note that depending on the atmospheric temperature, abundance estimates of some neutral elements may not be representative of the bulk abundance if they are heavily affected by ionization (e.g., Li, Na, K, Ca, Ba) or condensation (e.g., Ti, Sc, Al)~\citep{gibson_relative_2022, maguire_high-resolution_2023, maguire_high_2024, gandhi_retrieval_2023, pelletier_vanadium_2023}.

\begin{figure}[t]
\begin{center}
\includegraphics[width=\linewidth]{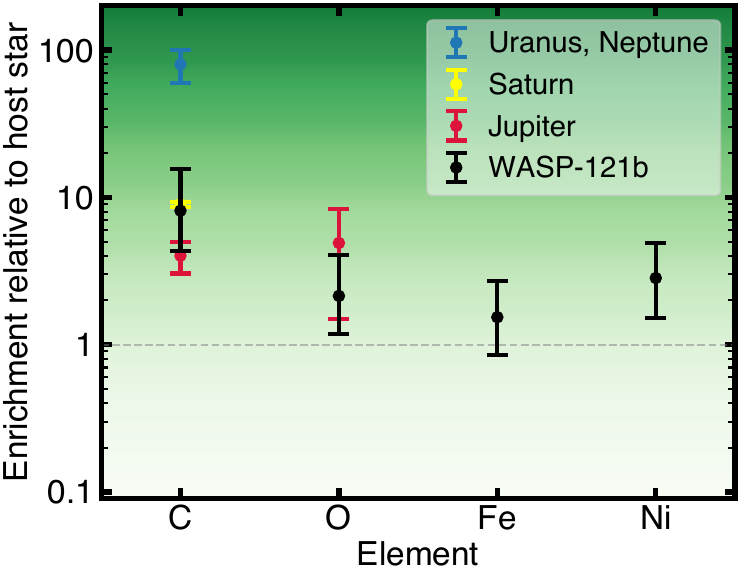}
\end{center}
\vspace{-4mm}
\caption{
Abundance enrichment for elements on WASP-121b compared to the Solar System giants.
Here the planetary atmosphere enrichment levels are measured relative to host star (gray dashed line) abundances, namely WASP-121~\citep{polanski_chemical_2022} for WASP-121b or the Sun~\citep{asplund_chemical_2021} otherwise.  
C abundances for the Solar System giants are inferred from CH$_4$ measurements (see \citet{atreya_deep_2020} and references therein) while the O abundance for Jupiter is from Juno~\citep{li_super-adiabatic_2024}.
No Fe or Ni measurements exist for Jupiter, Saturn, Uranus, or Neptune to compare to due to their cold temperatures causing all rock-forming elements to be condensed out of the upper atmosphere. 
While here all planets show enhanced metallicities relative to their host star for volatile elements (particularly C), WASP-121b does not show such an elevated enrichment for refractory elements.
}
\label{fig:abun_enrichment}
\end{figure}

If representative of the overall envelope, the markedly non-stellar elemental abundance ratios we measure on WASP-121b potentially suggests a different formation scenario than wide-orbit planetary mass companions, which are expected to retain abundance ratios closer to that of the native protoplanetary disc~\citep{hoch_assessing_2023}. 
Instead, our inferred atmospheric composition is more consistent with WASP-121b having accreted more ices than rocks over the course of its formation and evolution history.  The simultaneous measurement of volatile and refractory elements on a giant planet is also highly relevant in connection to our Solar System.
For Jupiter, Saturn, Uranus, and Neptune, we have abundance estimates of their volatile content but no direct measurements of any rock-forming elements.  In comparison, we find that WASP-121b, relative to its host star, has a degree of volatile (mainly C) enrichment that is broadly consistent with that of Jupiter, but does not have as high an enrichment for refractory elements Fe and Ni (Figure~\ref{fig:abun_enrichment}).
In the context of formation, this could mean that ice-rich environments in the protoplanetary disk are favorable for giant planets to form in.  However, we caution that this is only a single measurement that may not be representative of the overall giant planet population and that more measurements of different targets will be needed to draw any meaningful conclusions.

\subsection{Retrieval tests}  \label{subsec:retrieval_test}

In order to test the robustness of our results to both our data detrending procedure and some of the model assumption made, we perform a series of retrieval tests.

As the ESPRESSO observations cover a large portion of the full phase curve ($\phi$ roughly between 0.2 and 0.7, with $\phi$ = 0.5 corresponding to secondary eclipse), different facets of the hotter dayside and colder nightside of WASP-121b are visible at different orbital configurations.  This is in contrast to the CRIRES$^{+}$ observations which are more localized near secondary eclipse ($\phi$ roughly between 0.36 and 0.64) when the dayside of WASP-121b is expected to be most visible.  While we include a phase function to account for the overall change in flux observed at different orbital positions (Section~\ref{subsec:retrieval}, Step 4), it is possible that when furthest from $\phi$ = 0.5, the flux-weighted average chemistry and thermal structure observed are partly affected by contributions from the colder nightside regions of the planet and therefore differ from when the full dayside visible near eclipse.  
In order to test whether this could bias our results, we run a chemical equilibrium retrieval where we discard all ESPRESSO exposures more than 0.14 away from secondary eclipse in orbital phase (i.e., only using the subset of ESPRESSO observations that overlap with the CRIRES$^{+}$ data).  
We find that, other than slightly increasing the uncertainties on the derived parameters, this has almost no impact on the inferred composition of WASP-121b, with only the retrieved values for $K_p$ changing by more than 1$\sigma$ ($K_p = 216.49_{-0.30}^{+0.29}$\,km\,s$^{-1}$ for the main retrieval compared to $K_p = 217.03_{-0.32}^{+0.31}$\,km\,s$^{-1}$ for the phase-restricted retrieval). Interestingly, the higher inferred $K_p$ when considering a narrower portion of the orbit supports the idea that $K_p$ values measured from high-resolution emission spectroscopy of exoplanets are lower than the true Keplerian velocity as a result of the planetary rotation~\citep[][see their Figure 8]{hoeijmakers_mantis_2024}.

We also explore the possibility that the observed offset of the H$_2$O signal could result in an underprediction of its abundance and bias our inferred elemental abundance ratios.  For this, we re-ran a chemical equilibrium retrieval, but now including an additional radial velocity offset parameter applied on the H$_2$O cross-sections only.  We find that with this added degree of freedom, the model prefers to shift the H$_2$O opacity by $-2.38_{-1.04}^{+1.02}$\,km\,s$^{-1}$ relative to the overall signal from other species.  However, this only slightly changed the overall retrieved composition, resulting in a C/O ratio of $0.75_{-0.08}^{+0.06}$ compared to $0.73_{-0.08}^{+0.07}$ for the retrieval assuming no H$_2$O offset.

Finally, while our main analysis opted to remove 3 principal components to all observed times series', we also tested instead removing 5 PCs from each of the CRIRES$^{+}$ data sets and 10 PCs from each of the ESPRESSO data sets. Reassuringly, as it would be worrisome if the derived atmospheric properties of WASP-121b depended on the number of removed PCs, we found that this had very little impact on any of the retrieved parameters (all parameters consistent to less than 1$\sigma$ for both reductions). However, we note that derived atmospheric properties can depend on the number of PCs removed in some cases~\citep{smith_combined_2024}.

\subsection{Comparison with Lothringer et al.\ 2021}  \label{subsec:comparison}

While our results point towards WASP-121b having accreted an ice-rich envelope, this notably differs from a previous estimate based on broadband observations using the Hubble Space Telescope (STIS and WFC3)~\citep{lothringer_new_2021}. In that study, an ice-rich composition for the envelope of WASP-121b is ruled out based on a measured super-solar Fe/O, in apparent contradiction with our results.  While it is difficult to compare results obtained from space-based low-resolution spectrophotometry and ground-based high-resolution spectroscopy, we note some differences that may explain the different resulting conclusions.

In \cite{lothringer_new_2021}, the HST data is fit using the PETRA framework~\citep{lothringer_phoenix_2020}, which freely fits the H$_2$O, TiO, and VO abundances, with the remaining opacities from over 100 species calculated in chemical equilibrium parameterized as [Fe/H], a proxy for the overall metallicity.  This makes it less obvious what is driving the high retrieved [Fe/H], especially in combination with their extremely high retrieved VO abundance ($\sim$1000$\times$solar) and very low temperature (which is correlated with abundance).  In our case, we directly probe iron lines and thus our Fe/H is directly the iron-to-hydrogen ratio as opposed to the overall metallicity.  Our retrieved Fe abundance is also consistent with other studies probing the atmosphere of WASP-121b in transmission with ground-based high-resolution spectroscopy which find iron to be broadly consistent with a stellar composition~\citep{maguire_high-resolution_2023, gandhi_retrieval_2023}.  We also obtain similar results if using Ni (instead of Fe) as a tracer of refractories. On the flip side, it is possible that both Fe and Ni are under-abundant on WASP-121b (e.g., due to nightside condensation~\cite{ehrenreich_nightside_2020}), in which case probing these alone could result in an underestimation of the true refractory content.  However, other refractory species (Mg, V, Cr) that have different condensation temperatures in the atmosphere of WASP-121b also have stellar-like abundances~\citep{maguire_high-resolution_2023}.

We further note that our O/H abundance is primarily obtained from CO, which could not be constrained from the HST data in \cite{lothringer_new_2021} due to the limited sensitivity of STIS and WFC3 to CO opacity bandheads.  
The PETRA framework also fits the water abundance as being constant-with-altitude, which we found can bias the results towards a lower volatile content by not accounting for thermal dissociation (see also \cite{gandhi_revealing_2024, mansfield_metallicity_2024}). 
Finally, the HST observations are in transmission probing the terminator regions of the atmosphere while our CRIRES$^{+}$ and ESPRESSO data target the dayside thermal emission, making their direct comparison less obvious as these probe different regions of the atmosphere of WASP-121b. All in all, we caution that any comparison between these results should be made within the context of their significant differences.

\section{Link to planet formation} \label{sec:formation}

\begin{figure*}[t!]
\begin{center}
\includegraphics[width=\linewidth]{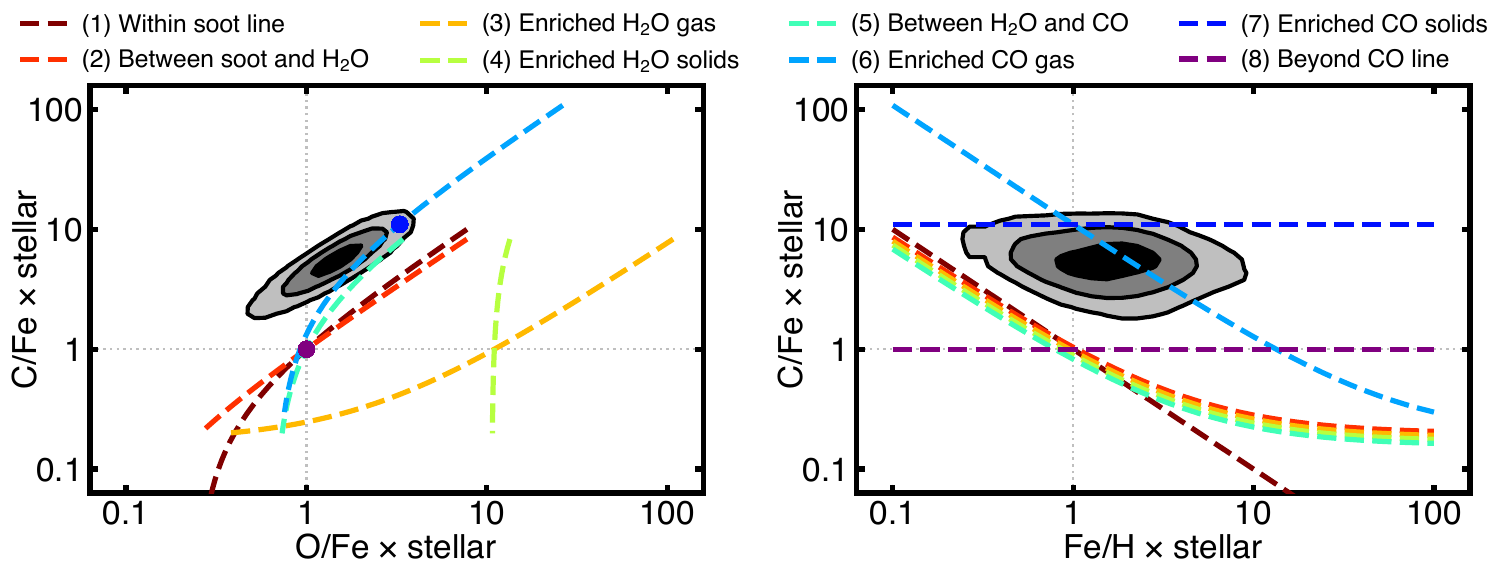}
\end{center}
\vspace{-4mm}
\caption{
Measured elemental abundance ratios on WASP-121b compared to predictions from different formation scenarios.
Here each colored dashed line corresponds to a predicted elemental composition for the considered giant planet formation scenarios (see \cite{chachan_breaking_2023} and Section~\ref{sec:formation}), compared to the retrieved values for WASP-121b (black and gray contours, 39.3\%, 86.5\%, and 98.9\% confidence intervals).  All abundance ratios are normalized to the stellar value (dotted gray lines).
Left panel:  O/Fe and C/Fe comparison, where the greater enrichment in C compared to O relative to Fe is consistent with formation scenarios beyond the H$_2$O snowline or near the CO snowline (scenarios 5, 6, and 7).
Right panel: Fe/H and C/Fe comparison now showing all scenarios except those enriched by CO gas or solids (scenarios 6 and 7, light and dark blue lines) being ruled out by the measured stellar-like refractory abundance.
Under this model, the only formation scenarios consistent with the measured refractory and volatile abundances is if WASP-121b accreted its envelope in a volatile-rich, high C/O environment near the CO snowline. 
}
\label{fig:chachan_figure}
\end{figure*}

In this section we explore potential formation scenarios that could give rise to the observed volatile-rich, high C/O atmosphere on WASP-121b, under the assumption that the measured composition is representative of the bulk envelope.
For this we use the protoplanetary disk model of \cite{chachan_breaking_2023} to predict the envelope composition or WASP-121b for formation scenarios at different orbital locations relative to various snowlines.
The framework takes as input a disk composition and outputs the planetary volatile and refractory elemental abundances for different envelope-accretion orbital positions.  
We calculate abundance ratios relative to refractory Fe, and scale everything with respect to the stellar abundances of WASP-121~\citep{polanski_chemical_2022}.
The framework considers eight formation scenarios at different orbital separations, which are briefly summarized here (see \cite{chachan_breaking_2023} for a full description).
\begin{itemize}
    \item Scenario 1: `Within soot line' considers formation in the inner hotter regions of the protoplanetary disk where any solid carbon will be sublimated out of the solid phase and into the gas phase.  In this region, accretion of large amounts of solids leads to carbon-poor compositions due to all the carbon being in vapor form.
    \item Scenario 2: `Between soot and H$_2$O' is past the soot line but before the H$_2$O snowline.  This is a similar environment to within the soot line, but with some carbon being held in solids. 
    \item Scenario 3: `Enriched H$_2$O gas' is just within the water snowline and considers the effect of pebble drift~\citep{oberg_excess_2016, booth_chemical_2017}.  As pebbles containing condensed H$_2$O ice drift inwards from farther out in the disk to now warmer regions, they can heat up and sublimate to locally enrich the gas phase O/H interior to the H$_2$O snowline.  
    \item Scenario 4: `Enriched H$_2$O solids' is just outside of the water snowline and rather considers the outward diffusion of enriched gas, which can then re-condense and enrich the H$_2$O content of solids slightly beyond the snowline~\citep{stevenson_rapid_1988}.
    \item Scenario 5: `Between H$_2$O and CO' is beyond the H$_2$O snowline but before the CO snowline.  In this region, H$_2$O is condensed as ice while CO remains in gas form.  Solids will therefore be O-rich, while the gas will have an elevated C/O ratio.
    \item Scenario 6: `Enriched CO gas' is analogous to scenario 3, but for CO instead of H$_2$O
    \item Scenario 7: `Enriched CO solids' is analogous to scenario 4, but for CO instead of H$_2$O
    \item Scenario 8: `Beyond CO line' considers the cold outer regions of the disk beyond the CO snowline where all C- and O-bearing compounds are frozen out of the gas phase.  Solids will therefore reflect the composition of the host star, meaning that the C/Fe and O/Fe ratios always remain constant regardless of the accreted solid-to-gas ratio.
\end{itemize}

The degree to which the local gas and solids will be enriched in scenarios 3, 4, 6, and 7 is assumed to be by a factor of ten in the model~\citep{chachan_breaking_2023}, although this can depend on the relative drift, sublimation, and diffusion timescales~\citep{booth_chemical_2017, aguichine_possible_2022, schneeberger_evolution_2023}.  
We note that all considered formation scenarios assume that envelope accretion occurs at a fixed location in the protoplanetary disk relative to snowlines. 
However, active migration during formation may cause the planet to drift across regions of the disk with different relative gas and solid compositions~\citep{turrini_tracing_2021}.  
In practice, the effect of such a mixing of material from various regions would manifest as an averaging of the predicted compositions for different scenarios.

The elemental abundance ratios predicted for the eight formation scenarios considered compared to our measured posteriors for the atmospheric composition of WASP-121b are shown in Figure~\ref{fig:chachan_figure}.  We find that the formation pathway that best explains the measured near-stellar Fe/H and super-stellar C/Fe and O/Fe is if WASP-121b accreted the bulk of its envelope from CO-enriched material (scenarios 6 or 7).   However, we notably cannot distinguish between whether the enhanced C content is the result of CO-enriched gas or solids as these predict overall similar abundance ratios in the case of a near-stellar amount of solids accreted.  Contrastingly, we can rule out formation scenarios near the H$_2$O snowline given that these would give rise to more O-rich, lower C/O ratio compositions. Overall our results indicate that WASP-121b could potentially have formed far out in the disc near the CO snowline more than a thousand times farther from its host star than where it is today~\citep{delrez_wasp-121_2016}. 

We note that for this comparison we use the elemental abundance ratios inferred from our hybrid free retrieval rather than our chemical equilibrium retrieval as these have wider, more conservative uncertainties (Table~\ref{tab:retrieval_results}).  However, similar conclusions are also reached if using the chemically consistent retrieval results.  We also verify our results in the case of using Ni instead of Fe as a tracer of the amount of refractories accreted by WASP-121b and obtain similar results, albeit with wider posteriors due to the weaker Ni signal.  Similar conclusions are also reached if we assume the protoplanetary nebula in which WASP-121b formed to have a solar composition~\citep{asplund_chemical_2021} rather than that of WASP-121~\citep{polanski_chemical_2022}.

To have formed in the outer protoplanetary disk and maintained a high fraction of ice relative to rock that is more comparable to model predictions for Uranus and Neptune~\citep{nettelmann_new_2013}, 
WASP-121b likely migrated to its present-day short orbital period without accreting significant amounts of refractory-rich planetesimals~\citep{lothringer_new_2021}.
Such a disk-free migration scenario is plausible given that WASP-121b lies on a polar orbit, perpendicular to the axis of rotation of its host star~\citep{delrez_wasp-121_2016, bourrier_hot_2020}.  A misaligned orbit could be indicative of a dynamical past, perhaps due to a planet-planet scattering event, the Kozai mechanism, and/or tidal circularization~\citep{nagasawa_formation_2008} that would have brought WASP-121b to its current orbital position without sweeping large amounts of rocky material concentrated near the disk midplane.

While we estimate the refractory content of WASP-121b via measurements of gaseous species present on its dayside atmosphere,
partial condensation may act as a sink for cloud-forming condensates thus making them appear under-abundant~\citep{parmentier_3d_2013, parmentier_transitions_2016}.  While the daysides of ultra-hot Jupiters are likely too hot to sustain a significant cloud mass, observations indicate~\citep{keating_uniformly_2019, ehrenreich_nightside_2020} and GCM models predict~\citep{parmentier_thermal_2018, helling_cloud_2021, komacek_patchy_2022} that condensation can still occur on the colder nightside.  While elements such as Fe cannot be fully condensed on WASP-121b based on the strong signals observed in both dayside emission and transmission~\citep[e.g.,][]{hoeijmakers_hot_2020, hoeijmakers_mantis_2024, gibson_detection_2020, merritt_inventory_2021}, it remains possible that some amount of refractories could be missing. 
In this case it is possible that our estimate of the Fe and Ni abundances underrepresent the true bulk refractory content of the atmosphere of WASP-121b.

Alternatively, the inferred composition enriched in volatiles relative to refractories could also be the result of inefficient ablation, i.e., the process of solids dissolving in the planetary envelope during accretion.  
While ice-rich bodies of sizes up to 1\,km are expected to be fully ablated in the outer envelopes of giant planets during accretion~\citep{pollack_formation_1996, iaroslavitz_atmospheric_2007, pinhas_efficiency_2016}, refractory planetesimals may not be fully dissolved before reaching the core, thus only partly contributing to enriching the atmosphere~\citep{venturini_planet_2016, brouwers_how_2018}.  
In this scenario, WASP-121b may still have a bulk super-stellar composition, but with an atmosphere that appears volatile-rich due to `missing' refractories locked away in the interior.
Such a process, however, would depend on 
the ability of the interior of a giant planet to retain material without it becoming well-mixed with the envelope~\citep{wahl_comparing_2017, vazan_jupiters_2018}.

Overall, if the processes that shaped WASP-121b are ubiquitous to giant planets, our results may indicate that environments with elevated ice-to-rock ratios are favorable for giant planet formation.  
In relation to our Solar System, it is therefore plausible that if Jupiter formed similarly to WASP-121b, it may also have volatiles enriched relative to refractories, as some estimates would indicate~\citep{bhattacharya_highly_2023}.

\section{Conclusions} \label{sec:conclusion}

We observed the dayside atmosphere of WASP-121b using the ESPRESSO and CRIRES$^{+}$ high resolution spectrographs. Via cross-correlation we detect strong emission signatures of volatile molecules CO and H$_2$O as well as refractory elements Fe and Ni near the expected orbital position of the planet. We find a slight velocity offset for H$_2$O, which is blueshifted by 2--3\,km\,s$^{-1}$ relative to the other species, potentially the result of an asymmetric distribution of H$_2$O across the dayside atmosphere of WASP-121b due to thermal dissociation.

To infer the temperature structure and elemental abundances of the dayside atmosphere of WASP-121b, we run high-resolution atmospheric retrievals. With both a modified `hybrid' free retrieval parameterizing H$_2$O dissociation and a chemical equilibrium retrieval fitting the overall volatile and refractory metallicities separately we find that WASP-121b has a refractory content consistent with its host star, but is comparatively enriched in volatiles.  Combined, we infer an elevated volatile-to-refractory ratio (a proxy of the ice-to-rock ratio) of $2.12_{-0.59}^{+0.77}$ $\times$ stellar as well as a super-stellar C/O = $0.87_{-0.06}^{+0.04}$ from our free retrieval.  We find overall consistent results from a chemical equilibrium retrieval, measuring a similar volatile-to-refractory enrichment of $1.75_{-0.41}^{+0.57}$ $\times$ stellar and a C/O ratio of $0.73_{-0.08}^{+0.07}$.  Our results therefore suggest that WASP-121b has an atmospheric enrichment in volatiles more akin to that of Jupiter, but no equivalently high enrichment for refractory species.

We further explore potential formation pathways for WASP-121b that could primordially produce such an ice-rich, high C/O atmosphere.
Applying the framework of \cite{chachan_breaking_2023} we find that the observed composition of WASP-121b is best matched if it would have accreted its envelope from CO-rich material near the CO snowline.  We highlight that measuring volatile-to-refractory ratios is an avenue where we can study the composition of giant planets that is otherwise not feasible for Jovian giants within in our Solar System.


\section{Acknowledgments}
This work is based on observations collected at the European Southern Observatory (ESO) under programmes 108.22CZ (PI:\,Allart, dPI:\,Pelletier), 0105.C-0591 (PI:\,Hoeijmakers), and 0106.C-0737 (PI:\,Hoeijmakers), using the CRIRES$^{+}$ and ESPRESSO spectrographs on the ESO VLT.  
CRIRES$^{+}$ is an ESO upgrade project carried out by Thüringer Landessternwarte Tautenburg, Georg-August Universität Göttingen, and Uppsala University. The project is funded by the Federal Ministry of Education and Research (Germany) through Grants 05A11MG3, 05A14MG4, 05A17MG2 and the Knut and Alice Wallenberg Foundation.  The authors acknowledge the ESPRESSO project team for its effort and dedication in building the ESPRESSO instrument.  This project has been carried out within the framework of the National Centre of Competence in Research PlanetS supported by the Swiss National Science Foundation under grant 51NF40\_205606.
We are grateful to the ESO staff members who made obtaining these observations possible.  We thank Thomas M.\ Evans-Soma for sharing the HST-retrieved temperature-pressure profile.
S.P.\ acknowledge the financial support from the Technologies for Exo-Planetary Science (TEPS) Natural Sciences and Engineering Research Council of Canada (NSERC) CREATE Trainee Program and the SNSF.
B.B.\ acknowledges funding by the NSERC, and the Fonds de Recherche du Québec -- Nature et Technologies (FRQNT).
R.A.\ is a Trottier Postdoctoral Fellow and acknowledges support from the Trottier Family Foundation. This work was supported in part through a grant from the FRQNT, and by the Institut Trottier de Recherche sur les Exoplanètes (iREx).
H.J.H.\ acknowledges contributions from eSSENCE (eSSENCE@LU 9:3) The Crafoord foundation and the Royal Fysiographic Society of Lund.
B.P.\ acknowledges partial financial support from The Fund of the Walter Gyllenberg Foundation.
P.S.\ acknowledges support provided by NASA through the NASA FINESST grant 80NSSC22K1598.
B.T.\ acknowledges the financial support from the Wenner-Gren Foundation (WGF2022-0041).

%

\vspace{5mm}
\facilities{VLT (CRIRES$^{+}$, ESPRESSO)}


\software{\texttt{Astropy}~\citep{astropy_collaboration_astropy_2013, astropy_collaboration_astropy_2018}, \texttt{NumPy}~\citep{harris_array_2020}, \texttt{SciPy}~\citep{virtanen_scipy_2020}, \texttt{scikit-learn}~\citep{pedregosa_scikit-learn_2011}, 
\texttt{Matplotlib}~\citep{hunter_matplotlib_2007}, \texttt{iPython}~\citep{perez_ipython_2007}, \texttt{emcee}~\citep{foreman-mackey_emcee_2013}, \texttt{corner}~\citep{foreman-mackey_cornerpy_2016}}






\bibliography{references}{}
\bibliographystyle{aasjournal}



\end{document}